\documentclass[a4paper]{article}

\usepackage[english]{babel}
\usepackage[utf8]{inputenc}
\usepackage{amstext,url,latexsym,amsfonts,amssymb,amsmath,amsthm}
\usepackage{graphicx}
\usepackage{xcolor}
\usepackage{tikz}
\usepackage{listings}
\usepackage{geometry}
\usepackage{fancyhdr}
\usepackage{microtype}
\usepackage[perpage]{footmisc}
\usepackage{float}
\usepackage{algorithm}
\usepackage{algorithmic}

\usetikzlibrary{shapes,snakes,patterns}
\tikzstyle{level 1}=[level distance=1.3cm, sibling distance=3.3cm]
\tikzstyle{level 2}=[level distance=1.3cm, sibling distance=1.6cm]
\tikzstyle{level 3}=[level distance=1.3cm, sibling distance=0.7cm]
\tikzstyle{level 4}=[level distance=1.3cm, sibling distance=0.4cm]
\tikzstyle{zero} = [circle,draw, inner sep=0pt, minimum size=2mm] 
\tikzstyle{un} = [circle,draw,fill=black,inner sep=0pt, minimum size=2mm]
\tikzstyle{unGraph}=[circle,draw,line width=2pt]
\tikzstyle{zeroGraph}=[circle,draw]

\pgfdeclarelayer{background}
\pgfsetlayers{background,main}
\tikzstyle{pointille} = [dashed,level distance=0.7cm]

\newcommand{\bydef}{\stackrel{\rm{def}}{=}}

\newcommand{\N}{\ensuremath{\mathbb{N}}}
\newcommand{\Sn}[1]{S(#1)}

\newtheorem{defi}{Definition}[section]
\newtheorem{lemma}{Lemma}[defi]
\newtheorem{prop}{Proposition}[defi]
\newtheorem{theorem}{Theorem}[section]

\title{Distributing labels on infinite trees}
\author{Nicolas Gast \and Bruno Gaujal}
\begin{document}
\maketitle
\begin{abstract}
  Sturmian words are infinite binary words with many equivalent
  definitions: They have a minimal factor complexity among all aperiodic
  sequences; they are balanced sequences (the labels 0 and 1 are as evenly
  distributed as possible) and they can be constructed using a mechanical
  definition.  All this properties make them good candidates for being
  extremal points in scheduling problems over two processors.  In this
  paper, we consider the problem of generalizing Sturmian words to
  trees. The problem is to evenly distribute labels 0 and 1 over infinite
  trees. We show that (strongly) balanced trees exist and can also be
  constructed using a mechanical process as long as the tree is irrational.
  Such trees also have a minimal factor complexity.  Therefore they bring
  the hope that extremal scheduling properties of Sturmian words can be
  extended to such trees, as least partially.  Such possible extensions are
  illustrated by one such example.
\end{abstract}


\section{Introduction}
In scheduling problems with an infinite number of tasks, the optimal
strategy  may  no longer be to execute tasks  ``as soon as
possible'' but rather ``as regularly as possible''.
Keeping this in mind, 
let us consider the following question: how to distribute ones and
zeros over an infinite sequence $w = (w_n)_{n \in \N}$ such that the ones (and the zeros) are
spread  as evenly as possible.
In a more formal way, the sequence $w$ is {\it balanced} if the number of
ones in a factor  $w_i,w_{i+1},\ldots,w_{i+\ell}$ of length $\ell$,  does not
vary by more than 1, for all $i$ and all $\ell$.
Such sequences exist and are called {\it Sturmian words}  when they
are not periodic.

Sturmian words are quite fascinating binary sequences: they have many
different characterizations formulated in terms coming from as many
mathematical frameworks, in which they always prove very useful.
For example, Sturmian words have a geometric description as digitalized straight
lines and as such have been used in computer visualization (see \cite{Klette}
for a review).
They can also be defined using an arithmetic characterization using a
repetitive rotation on a torus or continued fraction
decompositions. From  a combinatorial  point of
view, yet another characterization of Sturmian words is based on the
balance between ones and zeros in all factors, as mentioned before.
They are also used in symbolic dynamic system theory because they 
are aperiodic words with minimal factor complexity or because they
have palindromic properties.
Most of these equivalences have been known since the seminal work in
\cite{Morse}.

More recently, Sturmian sequences have also been used for optimization
purposes: they are extreme points of multimodular functions
\cite{Hajek,AGHBook}.
This has applications is scheduling theory . In \cite{GaujalH} rather
general scheduling problems with two processors are considered.
A simple case   is the following two processor mapping problem.
An infinite number of tasks of unit size are to be executed over two
processors (labeled $0$ and $1$) with related speeds, $v_0$ and $v_1$ such that $1/v_0 +
1/v_1 > 1$.
The tasks are released every time unit.
It is shown that an  optimal schedule (minimizing the average flow time) allocates
task $i$ to processor $w_i$ according to a sequence $w_1,w_2,w_2,\ldots $
that  is  Sturmian.

Another example solved in \cite{GaujalHVdL} is the following processor
allocation problem:
A single processor (with unit speed) is used to execute two types of tasks.
Tasks of type 1 (resp. 2) are released every time unit and are all of
size $S_0$ (resp. $S_1$). The allocation of the processor to the tasks
can be seen as a binary  sequence $w_1,w_2,\ldots$ saying which task is to be
served next.
Here also there  exists an  optimal Sturmian. sequence (minimizing the average flow time of all tasks).

Actually more general scheduling problems are solved by Sturmian
sequences. For instance of the tasks are released according to a
stationary process and the task sizes are also stochastic, independent
of the release process, then both problems mentioned above are also 
solved by Sturmian sequences.

A natural extension is to consider the case where more than two
processors can be used to execute the tasks.
This leads to the construction of generalized  Sturmian words in
several direction.

The first one  is to study words using more than two
letters. Billiard sequences
in hypercubes extent the torus definition of Sturmian sequences while episturmian sequences
\cite{berstel} extend  the palindromic characterization of Sturmian
words. Unfortunately, both extensions differ substantially and none of
them provides an optimal schedule for the $k$ processor mapping problem.

Another extension is to two dimensions. A complete characterization of
two-dimensional non-periodic sequences with minimal complexity is
given in \cite{Cassaigne}. Here again  the alternative
characterizations are lost.

Yet another generalization is to trees \cite{berstel2007fis}, where Sturmian
trees are defined as infinite binary automata  such that the number of
factors (sub-trees) of size $n$ is $n+1$.
The other characterizations of Sturmian words are
lost once more.

Finally, another extension of Sturmian  concerns discrete
planes. Here, several 
characterizations  of Sturmian lines can be extended to discrete
planes. Interesting relations 
between multidimensional continued fraction decomposition of the
normal direction of the plane and the patterns of its discretization
mimic what happens for Sturmian sequences,  \cite{Fernique}.

The aim of this paper is to do the same for trees.
We  introduced in \cite{GastGaujal}  a new type of infinite 
trees: unordered trees, for which the left and right children of each
node are not distinguishable and  gave a brief presentation of its
main properties.
Here, We make an exhaustive study  of  such trees. We show that the balance property (distributing evenly
the labels equal to one or zero over the vertex of the tree)
coincides with a characterization of trees using integer parts of
affine functions (called mechanicity). 
Furthermore these balanced trees have a minimal factor complexity.
Therefore, they can be seen as a natural extension of Sturmian 
sequence in more than one aspect. This brings some  hope to use them as extreme points for  adapted
optimization problems.

Our purpose in the paper is two-fold.  The first part of the paper is
dedicated to the study of general unordered infinite trees with binary labels.  We
provide definitions of the main concepts as well as the basic properties of
unordered trees with a special focus on the notion of density (the average
number of ones) and rationality  The second part of the paper investigates balanced
unordered trees and their properties. In particular we show that strongly
balanced trees (defined later) are mechanical (so that they have a density
and all labels can be constructed in almost constant time). Furthermore
their factor complexity is minimal among all non-periodic trees.

We also investigate rational balanced trees by showing that their
density is easy to compute and by providing an algorithm with
polynomial complexity to test whether a rational tree is strongly
balanced. Finally, we show that balanced trees are extremal points of some 
convex functions, bringing some hope that they can be used to solve
optimization problems.

\section{Infinite Trees}

For {\it ordered infinite trees} ,  we follow the presentation given
in \cite{berstel2007fis}.
Ordered infinite trees are automata with an infinite number of states.
An automata is a tree-automaton  if it  has one initial state and each state has a uniform
in-degree equal to one (except for the initial state, whose in-degree
is 0) and a uniform out-degree $d$ with  labels $a_1,\cdots, a_d$ on the arcs.
Every node $v$ is labeled  by $\ell(v)=1$ (resp. $0$) if it is final (resp. non-final).

The language accepted by the tree-automaton  ${\cal T}$ is a
subset of ${\cal A}^* $ (where the alphabet ${\cal A} =  \{a_1, \ldots
a_d\}$) and is denoted by   ${\cal L}({\cal T})$.
Thus, a word  $w$ in the free monoid ${\cal A}^*$   corresponds to a
node in ${\cal T}$,
and a   word  $w$ in  ${\cal L}({\cal T})$ corresponds to a node in ${\cal T}$
with label $1$.
Conversely, a unique tree-automaton  can be associated to any subset $L$ of
${\cal A}^* $,
by labeling by one the  nodes corresponding to the words in $L$.

Classically for automata, a family of equivalence relations can be defined
over the nodes of tree ${\cal T}$:
$v \sim_0 u$ if $\ell(v) = \ell(u)$,
$v\sim_{n+1} u $ if $v\sim_{n} u $ and for all $i$, the $i$th child of $u$,
$ua_i $ and the $i$th child of $v$, $va_i$ satisfy
$ua_i \sim_{n} va_i$.
By definition of $\sim_n$, $u \sim_n v$ if and only if the subtree rooted in $u$ of
height $n$ is the same  as the subtree  rooted in $v$ of
height $n$.

${\cal L({\cal T})}$ is recognized by its minimal deterministic automaton (possibly infinite), say $A({\cal T})$.
Actually, $A({\cal T})$   can be obtained from the tree ${\cal T}$ by
merging all the states in the tree in the same equivalence
  classes of $\sim_n$ for all $n$.

\begin{figure}[hbtp]
  \centering
  \begin{tabular}{cc}
    \begin{tabular}{c}
      \begin{tikzpicture}[yscale=0.6,xscale=0.8]
  \newcommand\aaa{}
  \newcommand\bbb{}
  \newcommand\aaaa{}
  \newcommand\bbbb{}
  
  \node[un] {} 
  child   {node[un] {}
    child {node[zero] {} 
      child {node[zero] {} 
        child {node[zero] {} {child[pointille]{}}
          edge from parent
          node[above,pos=.9]{\aaaa}}
        child {node[zero] {} {child[pointille]{}}
          edge from parent
          node[above,pos=.9]{\bbbb}}
        edge from parent
        node[above,pos=.8] {\aaa}}
      child {node[zero] {} 
        child {node[zero] {} {child[pointille]{}} edge from parent node[above,pos=.9]{\aaaa}}
        child {node[zero] {} {child[pointille]{}} edge from parent node[above,pos=.9]{\bbbb}}
        edge from parent
        node[above,pos=.8] {\bbb}}
      edge from parent
      node[above,pos=.8] {$a$}}
    child {node[un] {} 
      child {node[un] {} 
        child {node[un] {}
          child {node [zero] {}  {child[pointille]{}}edge from parent node[above,pos=.9]{\aaaa}}
          child {node [un] {}
            child {node [un] {}  {child[pointille]{}}edge from parent node[above,pos=.9]{\aaaa}}
            child {node [zero] {} {child[pointille]{}
                edge from parent node[above,pos=.9]{\bbbb}}}
            edge from parent node[above,pos=.9]{\bbbb}}
          edge from parent node[above,pos=.9]{\aaaa}}
        child {node[zero] {} {child[pointille]{}} edge from parent node[above,pos=.9]{\bbbb}}
        edge from parent
        node[above,pos=.8] {\aaa}}
      child {node[zero] {}
        child {node[zero] {} {child[pointille]{}} edge from parent node[above,pos=.9]{\aaaa}}
        child {node[zero] {} {child[pointille]{}} edge from parent node[above,pos=.9]{\bbbb}}
        edge from parent
        node[above,pos=.8] {\bbb}}
      edge from parent
      node[above,pos=.8] {$b$}}
    edge from parent
    node[above,pos=.5] {$a$}}
  child {node[zero] {}
    child {node[zero] {}
      child {node[zero] {}
        child {node[zero] {} {child[pointille]{}} edge from parent node[above,pos=.9]{\aaaa}}
        child {node[zero] {} {child[pointille]{}} edge from parent node[above,pos=.9]{\bbbb}}
        edge from parent
        node[above,pos=.8] {\aaa}}
      child {node[zero] {} 
        child {node[zero] {} {child[pointille]{}} edge from parent node[above,pos=.9]{\aaaa}}
        child {node[zero] {} {child[pointille]{}} edge from parent node[above,pos=.9]{\bbbb}}
        edge from parent
        node[above,pos=.8] {\bbb}}
      edge from parent
      node[above,pos=.8] {$a$}}
    child {node[zero] {} 
      child {node[zero] {} 
        child {node[zero] {} {child[pointille]{}} edge from parent node[above,pos=.9]{\aaaa}}
        child {node[zero] {} {child[pointille]{}} edge from parent node[above,pos=.9]{\bbbb}}
        edge from parent
        node[above,pos=.8] {\aaa}}
      child {node[zero] {}
        child {node[zero] {} {child[pointille]{}} edge from parent node[above,pos=.9]{\aaaa}}
        child {node[zero] {} {child[pointille]{}} edge from parent node[above,pos=.9]{\bbbb}}
        edge from parent
        node[above,pos=.8] {\bbb}}
      edge from parent
      node[above,pos=.8] {$b$}}
    edge from parent
    node[above,pos=.5] {$b$}};
\end{tikzpicture}
    \end{tabular}
    &
    \begin{tabular}{c}
      \begin{tikzpicture}[auto,node distance=1.3cm,semithick,shorten >=1pt,yscale=0.6,xscale=0.8]

  \foreach \x/\y in {0,...,6}
  \node[unGraph] (\y) at (1.3*\y,0) {\y};
    
  \node (ENTREE) at(0,1.5) {};
  \node[zeroGraph] (END) at (3*1.3,-3) {$\infty$};
  \node[circle,pointille] (7) at (1.3*7,0) {...};
  
  \newcounter{y}
  \foreach \z/\y/\x in {a/0/1,b/1/2,a/2/3,a/3/4,b/4/5,a/5/6}
  \draw (\y) edge[->] node{\z} (\x)
  (\y) edge[->] 
  (END);
  
  \draw (6) edge[->,pointille] node{b} (7)
  (6) edge[->] (END)
  (END) edge[loop below,->] node{a,b} ()
  (ENTREE) edge[->] (0);

  \foreach \x/\y in {b/0,a/1,b/2,b/3,a/4,b/5,a/6}
  \node[fill=white,circle,inner sep=1pt] at (1.7+0.735*\y,-1.4) {\x};
\end{tikzpicture}

    \end{tabular}
  \end{tabular}
  \caption{The tree-automaton  recognizing the Fibonacci word  and the
    corresponding  minimal automaton}
  \label{fig:fibo}
\end{figure}
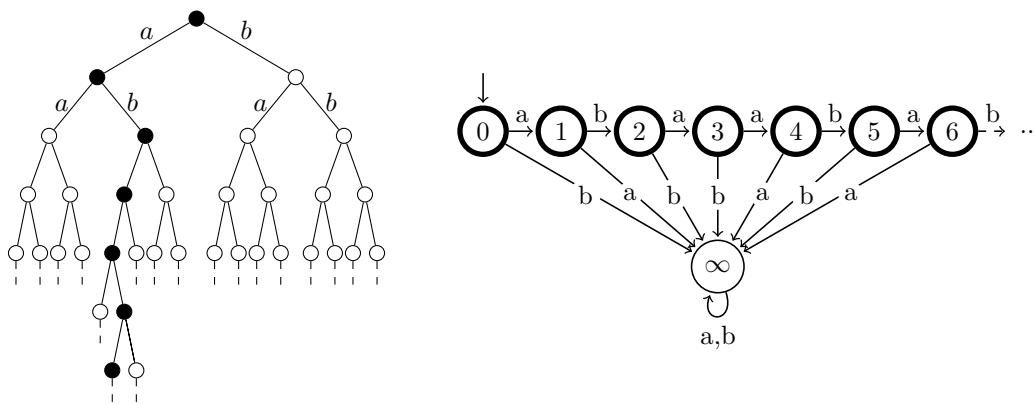

An example is given in Figure \ref{fig:fibo} where the 
infinite tree-automaton  and the minimal automaton  recognizing all the prefixes of
the Fibonacci word (over the alphabet $\{ a,b\}$) is given together with
the corresponding minimal automaton (which has an infinite number of states).

The number of subtrees  of size $k$ in ${\cal T}$ is called the complexity
$P(k)$, of ${\cal T}$. $P(k)$ is the number of equivalence classes of $\sim_k$.
If $P(k) \leq k$ for at least one $k$, then it can be shown
(\cite{berstel2007fis}) that the   complexity  is bounded by $k$. This implies that  the minimal automaton $A({\cal T})$ has $k$ states. The tree is therefore
called rational, since it recognizes a rational language.

If a tree-automaton  ${\cal T}$ is such that $P(k) = k+1$ for all $k$, then it has a minimal complexity
among all non-rational trees. Such trees have been shown to exist and are called Sturmian
in \cite{berstel2007fis} by analogy with the factor complexity definition of Sturmian
words. In \cite{berstel2007fis} 
several classes of Sturmian tree-automata  are presented.
However such trees are not balanced and cannot be defined using a
mechanical construction, as with Sturmian words. \\

In the following  we rather consider a different type of trees, 
namely infinite directed {\it graphs}  with labels $0$ or $1$ on nodes and
with uniform in-degree $1$ and out-degree $d \geq 2$. Here,  the children 
of a node are not ordered.  Thus, the  main difference with the previous definition is that arcs
are not labeled.
Therefore such trees cannot be bijectively associated with languages. 
However, it is possible to construct a minimal multi-graph ({\it i..e.} with
multiples arcs) $G(T)$ associated
with the tree $T$, mimicking the construction of the minimal automaton for
ordered trees.
Let us consider a family of equivalence relations over the nodes of
$T$:\\
$ v \equiv_0 u$  if $u$ and $v$ have the same label: $\ell (u ) = \ell (v)$
and\\
$v \equiv_{n+1} u $ if $ v \equiv_{n} u$ and the children of $v$ are 
equivalent (for  $\equiv_{n}$) to the children of $u$.\\
Therefore,  $ v \equiv_{n} u$ if and only if the subtree with root $v$
of height  $n$ is isomorphic to the subtree with root $u$ with height  $n$.
By merging the nodes of $T$ when they belong to the same equivalence
classes, for all $n$, one gets the minimal multi-graph $G(T)$ of the factors of
$T$: all nodes merged in the same vertex of $G(T)$ have the same
subtrees of every height.

In $G(T)$, the node corresponding to the root of $T$ is distinguished.
(graphically, this is done by adding an arrow pointing to the node).

There exists a way to associate an ordered tree-automaton  ${\cal T}$ to a  tree $T$ by
choosing an order on the children of each node.
This can be done by seeing $G(T)$ as an automaton 
by labeling arcs in $G(T)$ with letters $a_1,\ldots a_d$ in an arbitrary
fashion.
Conversely, a tree-automaton ${\cal T}$ can be converted into a graph
$T$ 
by removing the labels on the arcs. This graph is called the
unordered version of ${\cal T}$.

\begin{figure}[hbtp]
  \centering
  \begin{tabular}{cc}
    \begin{tabular}{c}
      \begin{tikzpicture}[yscale=0.6]
  \node[un] {} 
  child {node[zero] {}
    child {node[zero] {}
      child {node[zero] {}
        child {node[zero] {} {child[pointille]{}}}
        child {node[zero] {} {child[pointille]{}}}}
      child {node[zero] {} 
        child {node[zero] {} {child[pointille]{}}}
        child {node[zero] {} {child[pointille]{}}}}}
    child {node[zero] {} 
      child {node[zero] {} 
        child {node[zero] {} {child[pointille]{}}}
        child {node[zero] {} {child[pointille]{}}}}
      child {node[zero] {}
        child {node[zero] {} {child[pointille]{}}}
        child {node[zero] {} {child[pointille]{}}}}}}
  child {node[un] {} 
    child {node[zero] {} 
      child {node[zero] {} 
        child {node[zero] {} {child[pointille]{}}}
        child {node[zero] {} {child[pointille]{}}}}
      child {node[zero] {} 
        child {node[zero] {} {child[pointille]{}}}
        child {node[zero] {} {child[pointille]{}}}}}
    child {node[un] {} 
      child {node[zero] {}
        child {node[zero] {} {child[pointille]{}}}
        child {node[zero] {} {child[pointille]{}}}}
      child {node[un] {} 
        child {node[zero] {} {child[pointille]{}}}
        child {node[un] {}
          child {node [zero] {}  {child[pointille]{}}}
          child {node [un] {}
            child {node [zero] {} {child[pointille]{}}}
            child {node [un] {}  {child[pointille]{}}}}}}}};

\end{tikzpicture}
    \end{tabular}
    &
    \begin{tabular}{c}
      \begin{tikzpicture}[auto,node distance=1.3cm,semithick,shorten >=1pt,yscale=0.6]
  
  \tikzstyle{noeud}=[circle,draw];

  \node[noeud,line width=2pt] (0) at (0,0) {0};
  \node (ENTREE) [left of=0] {};
  \node[noeud] (END) at (0,-3) {$\infty$};
  
  \draw (0) edge[loop right,->] (0)
  (0) edge[->] (END)
  (END) edge[loop right,->] ()
  (END) edge[out=20,in=-20,loop,->] ()
  (ENTREE) edge[->] (0);
  
\end{tikzpicture}

    \end{tabular}
  \end{tabular}
  \caption{A tree and the associated minimal multi-graph.}
  \label{fig:fibo_unordered}
\end{figure}
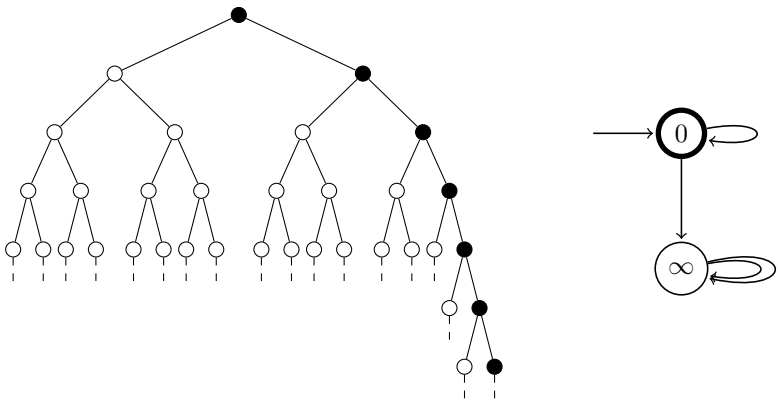

An example of an unordered tree is given in Figure
\ref{fig:fibo_unordered}.
The  label of the black (white)  node is 1  (0). The arcs are
implicitly directed from top to bottom.
Actually, most figures in this paper will represent binary trees (with
out-degree $d = 2$), although all the discussion is carried throughout
for arbitrary degrees.
The nodes of the  associated multi-graph $G(T)$  are numbered arbitrarily and
nodes with label 1 are displayed with a bold circle. The node
corresponding to the root of the tree is pointed by an arrow.
This tree can be seen as the tree-automata recognizing the Fibonacci
word where the labels on the arcs have been removed (there is no
longer a  left and right child at each node).
Note that while the minimal automaton is infinite (see Figure
\ref{fig:fibo}),
the minimal graph $G(T)$ is finite, with two nodes, one correspond to
the tree where all labels are $0$ and one with all labels equal to 0
expect on one  branch (see Figure \ref{fig:fibo_unordered}).

\subsection{Irreducibility and periodicity}

By analogy with Markov chains, a tree $T$ is {\it irreducible} if $G(T)$ is strongly
connected.
Also, an irreducible tree $T$ is {\it periodic} with period $p$ if
the greatest common divisor 
of the lengths of all 
cycles in $G(T)$ is $p$.
A tree with period 1 is also called {\it aperiodic}.

\subsection{Factors, complexity and Sturmian trees}

A {\it factor of size $n$ (and width $1$)} is a subgraph of $T$ which is a
complete subtree of height $n$.
The number of nodes in a factor of size $n$ is denoted by $\Sn{n} \bydef \frac{d^n-1}{d-1}$.

A {\it factor of size   $n$ and width $k$} (with root $v$), 
is a sub-graph of $T$ which is the subtree of height  $k+n$ rooted in
$v$  minus the subtree of height  $k$, rooted in $v$.
The number of nodes  of a factor of size  $n$ and width $k$ is
 $\Sn{n,k} \bydef \frac{d^{n+k} - d^k}{d-1}$.

Similarly to what as been done  for words, the factor {\it complexity}
${\cal P}_T(n)$
of a tree $T$ is the number of distinct  factors of size $n$ and width
1. 

The complexity of a tree ${\cal P}_T(n)$ can be bounded by the total
number of  ways to label trees of height $n$ and degree $d$, say $A_n$. 

It should be clear that $A_1 = 2$ (a node can be labeled 0 or 1)
and that $A_{n+1} = 2 M(A_{n},d)$ where $M(x,y)$ is the number of
multisets with $y$ elements taken from a set with $x$ elements.
Therefore using binomial coefficients,
\[  A_{n+1} = 2  \binom{A_n + d -1}{A_n-1}.\]
This is a polynomial recurrence equation of degree $d$.
A change of variable, $u_n = \log A_n + \frac{1}{d-1}
\log\frac{2}{d!}$ yields a new recurrence equation $u_{n+1} = d u_n +
\varepsilon_n$ where $\epsilon_n = o(1)$.
This implies that $A_n = \phi^{d^n+ o(d^n)}$ for some $\phi$  with $1 < \phi <
2$.

As for lower bounds on the complexity of a tree,
it will be shown in Section \ref{sec:rat} that 
trees such that ${\cal P}_T(n) \leq n$  for at least one $n$ are
rational,
{\it i.e.} have a bounded number of factors of any size (this means
that the minimal multi-graph is finite).

Therefore, trees $T$ such that $G(T)$ is infinite and with a minimal
complexity
should satisfy ${\cal P}_T(n) = n+1$.
These trees will be  called {\it Sturmian trees} by analogy with words.
It is not difficult to exhibit such trees. For example, starting with
a Sturmian word $w$  a binary tree such
that  all nodes on level $i$ have label $w_i$ is Sturmian.

\begin{figure}[ht]
  \centering
  \begin{tabular}{cc}
    \begin{tabular}{c}
      \begin{tikzpicture}[yscale=0.6]
  \node[un]{}
  child {node[zero]{}
    child {node[zero]{}
      child {node[zero]{}
        child {node[zero]{} child[pointille] }
        child {node[zero]{} child[pointille] }}
      child {node[zero]{}
        child {node[zero]{} child[pointille] }
        child {node[un]{} child[pointille] }}}
    child {node[un]{}
      child {node[zero]{}
        child {node[zero]{} child[pointille] }
        child {node[un]{} child[pointille] }}
      child {node[zero]{}
        child {node[zero]{} child[pointille] }
        child {node[zero]{} child[pointille] }}}}
  child {node[zero]{}
    child {node[zero]{}
      child {node[zero]{}
        child {node[zero]{} child[pointille] }
        child {node[zero]{} child[pointille] }}
      child {node[zero]{}
        child {node[zero]{} child[pointille] }
        child {node[zero]{} child[pointille] }}}
    child {node[zero]{}
      child {node[zero]{}
        child {node[zero]{} child[pointille] }
        child {node[zero]{} child[pointille] }}
      child {node[zero]{}
        child {node[zero]{} child[pointille] }
        child {node[zero]{} child[pointille] }}}};
\end{tikzpicture}
  
    \end{tabular}
    &
    \begin{tabular}{c}
      \begin{tikzpicture}[auto,node distance=1.3cm,semithick,shorten >=1pt,yscale=0.6,xscale=0.7]

  \tikzstyle{noeud}=[circle,draw];
  \foreach \x/\y in {1,...,5}
  \node[noeud] (\y) at (1.3*\y,0) {\y};
  
  \node[noeud,line width=2pt] (0) at (0,0) {0};
  
  \node (ENTREE) at(0,1.5) {};
  \node[noeud] (END) at (-1.3,0) {$\infty$};
  \node[circle,pointille] (6) at (1.3*6,0) {...};
  
  \foreach \x/\y in {0/1,1/2,2/3,3/4,4/5,5/6}
  \draw (\y) edge[->,bend left] (\x)
  (\x) edge[->,bend left] (\y);
  
  \draw (END) edge[loop below,->]()
  (END) edge[loop above,->]()
  (ENTREE) edge[->] (0)
  (0) edge[->] (END);

\end{tikzpicture}

    \end{tabular}
  \end{tabular}
  \caption{The Dyck tree and its minimal graph.}
  \label{fig:dyck_tree}
\end{figure}
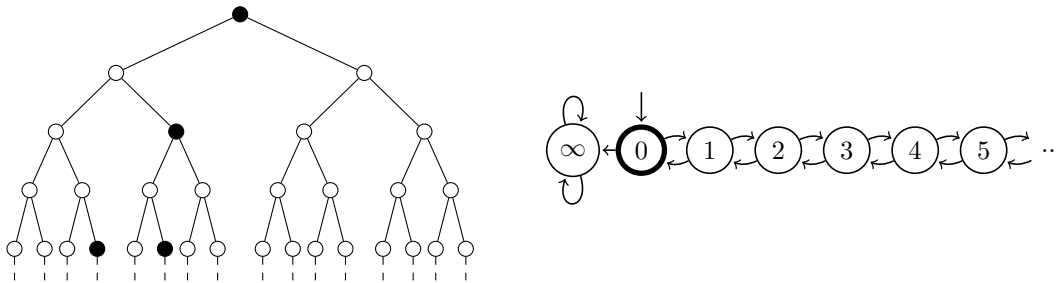

Another more interesting example is the Dyck tree.
The Dyck tree is 
  represented on Figure \ref{fig:dyck_tree}. This tree is the
  unordered version of the tree-automata recognizing 
  the Dyck language (language generated by the context-free grammar $S\to
  aSbS|\epsilon$) and it is not hard to see that this tree is Sturmian.
For that,  consider the graph $G(T)$ associated with the Dyck tree $T$,
also displayed in Figure \ref{fig:dyck_tree}.

There are two factors of size $1$ in $T$: those with a root labeled 1 (all
associated with node 0 in $G(T)$) and those with a root labeled 0
(associated with nodes $\infty, 1,2,\cdots$ in $G(T)$).
This corresponds to the equivalence classes for $\equiv_1$.

As for factors of size $n$,
all those with a root in node $\infty$ and $n,n+1,n+2 $ have all their
labels equal to 0: no  path of length $n$ in $G(T)$  reaches  the
only node with label 1, namely node 0.

As for the factors starting in node $i$ of $G(T)$ with $0\leq i < n$,
then the first node with label $1$ is at level $i+1$.
This means that all these factors are distinct.
In other words, the equivalence classes for $\equiv_n$ are
$\{\infty, n,n+1,\ldots\} , \{0\},\{1\},\ldots, \{n-1\}$.
The number of distinct factors of size $n$ is therefore
$n+1$.

\subsection{Density}

The density of a tree $T$ is meant to capture the average number of $1$ in the
tree.

For a node $v$ and a height $n\geq 0$, we define the density of the factors
of size $n$ with root $v$ by the average number of nodes with label $1$ in
this sub-tree. 
Let us call $d_v(n)$ the density of the factor of size $n$with root $v$ and 
let $r$ be the root of the tree $T$.
In the following we will be using four notions of density.

\begin{itemize}
\item The {\it rooted density}  of the tree is the limit of the density of the
  sub-trees  of the root $r$ (if it exists): 
  \[ \lim_{n\to\infty} d_r(n)\]
\item The {\it rooted average density} of the tree the Cesaro limit of these
  densities: 
  \[ \lim_{n\to\infty} \frac{1}{n}\sum_{i=1}^{n}d_r(n) \]
\item The  {\it density}  of the tree is $\alpha$ if it has an
  identical  rooted density for all node $v$: 
  \[ \forall v: \alpha = \lim_{n\to\infty} d_v(n)\]
\item The {\it average density} of the tree is $\alpha$ if it has an
  identical rooted
  average density for all node $v$: 
  \[ \forall v: \alpha = \lim_{n\to\infty} \frac{1}{n}\sum_{i=1}^{n} d_v(n)\] 
\end{itemize}

From the definition, we have the following direct implications:
If a tree admits a density, then  it admits an average density. In turn,
a tree with a average density also has 
a rooted average density.
Also,  a tree with a density has a rooted density.

Although the rooted definitions seem more natural and simple ,
the  definition of  general densities have  the advantage
that they do  depend on the choice of the
root. See Figure \ref{fig:having_a_density} for some examples.
These examples will be further developed in the following section on
rational trees.

\begin{figure}[ht]
  \centering
  \begin{tabular}{ccc}
    \begin{tikzpicture}[xscale=0.6, yscale=0.6]
  \node[zero] {} 
  child {node[un] {} 
    child {node[un] {} 
      child {node[un] {} 
        child {node[un] {} {child[pointille]{}}}
        child {node[zero] {} {child[pointille]{}}}}
      child {node[zero] {} 
        child {node[un] {} {child[pointille]{}}}
        child {node[zero] {} {child[pointille]{}}}}}
    child {node[zero] {} 
      child {node[un] {} 
        child {node[un] {} {child[pointille]{}}}
        child {node[zero] {} {child[pointille]{}}}}
      child {node[zero] {}
        child {node[un] {} {child[pointille]{}}}
        child {node[zero] {} {child[pointille]{}}}}}}
  child {node[zero] {}
    child {node[un] {}
      child {node[un] {}
        child {node[un] {} {child[pointille]{}}}
        child {node[zero] {} {child[pointille]{}}}}
      child {node[zero] {} 
        child {node[un] {} {child[pointille]{}}}
        child {node[zero] {} {child[pointille]{}}}}}
    child {node[zero] {} 
      child {node[un] {} 
        child {node[un] {} {child[pointille]{}}}
        child {node[zero] {} {child[pointille]{}}}}
      child {node[zero] {}
        child {node[un] {} {child[pointille]{}}}
        child {node[zero] {} {child[pointille]{}}}}}};
\end{tikzpicture}&
    \begin{tikzpicture}[xscale=0.6, yscale=0.6]
  \node[zero] {} 
  child {node[un] {} 
    child {node[zero] {} 
      child {node[un] {} 
        child {node[zero] {} {child[pointille]{}}}
        child {node[zero] {} {child[pointille]{}}}}
      child {node[un] {} 
        child {node[zero] {} {child[pointille]{}}}
        child {node[zero] {} {child[pointille]{}}}}}
    child {node[zero] {} 
      child {node[un] {} 
        child {node[zero] {} {child[pointille]{}}}
        child {node[zero] {} {child[pointille]{}}}}
      child {node[un] {} 
        child {node[zero] {} {child[pointille]{}}}
        child {node[zero] {} {child[pointille]{}}}}}}
  child {node[un] {} 
    child {node[zero] {} 
      child {node[un] {} 
        child {node[zero] {} {child[pointille]{}}}
        child {node[zero] {} {child[pointille]{}}}}
      child {node[un] {} 
        child {node[zero] {} {child[pointille]{}}}
        child {node[zero] {} {child[pointille]{}}}}}
    child {node[zero] {} 
      child {node[un] {} 
        child {node[zero] {} {child[pointille]{}}}
        child {node[zero] {} {child[pointille]{}}}}
      child {node[un] {} 
        child {node[zero] {} {child[pointille]{}}}
        child {node[zero] {} {child[pointille]{}}}}}};
\end{tikzpicture}&
    \begin{tikzpicture}[xscale=0.6, yscale=0.6]
  \node[zero] {} 
  child {node[un] {} 
    child {node[un] {} 
      child {node[un] {} 
        child {node[un] {} {child[pointille]{}}}
        child {node[un] {} {child[pointille]{}}}}
      child {node[un] {} 
        child {node[un] {} {child[pointille]{}}}
        child {node[un] {} {child[pointille]{}}}}}
    child {node[un] {} 
      child {node[un] {} 
        child {node[un] {} {child[pointille]{}}}
        child {node[un] {} {child[pointille]{}}}}
      child {node[un] {}
        child {node[un] {} {child[pointille]{}}}
        child {node[un] {} {child[pointille]{}}}}}}
  child {node[zero] {} 
    child {node[zero] {} 
      child {node[zero] {} 
        child {node[zero] {} {child[pointille]{}}}
        child {node[zero] {} {child[pointille]{}}}}
      child {node[zero] {} 
        child {node[zero] {} {child[pointille]{}}}
        child {node[zero] {} {child[pointille]{}}}}}
    child {node[zero] {} 
      child {node[zero] {} 
        child {node[zero] {} {child[pointille]{}}}
        child {node[zero] {} {child[pointille]{}}}}
      child {node[zero] {}
        child {node[zero] {} {child[pointille]{}}}
        child {node[zero] {} {child[pointille]{}}}}}};
\end{tikzpicture}
  \end{tabular}
  \caption{The first  tree has a density of $1/2$, the second one an
    average density equal to $1/2$ but no density. The last one has a
    rooted density 1/2  but no average density.}
  \label{fig:having_a_density}
\end{figure}

\section{Rational trees  \label{sec:rat}}

A tree $T$ is {\it rational } if the associated minimal multi-graph
$G(T)$ is finite.

An example of a  rational tree $T$  is displayed in Figure \ref{fig:rat}
together with its graph $G(T)$.
Note that this tree is not irreducible. One final strongly component of
period 2 (it corresponds to the alternating subtrees starting with ones and
zeros displayed on the left) while the other one is aperiodic (it
corresponds to the subtree with all its labels equal to one, displayed on
the right).

\begin{figure}[hbtp]
  \centering
  \begin{tabular}{cc}
    \begin{tabular}{c}
      \begin{tikzpicture}[yscale=0.6]
  \node[zero] {} 
  child {node[zero] {} 
    child {node[un] {} 
      child {node[zero] {} 
        child {node[un] {} {child[pointille]{}}}
        child {node[un] {} {child[pointille]{}}}}
      child {node[zero] {} 
        child {node[un] {} {child[pointille]{}}}
        child {node[un] {} {child[pointille]{}}}}}
    child {node[un] {} 
      child {node[zero] {}
        child {node[un] {} {child[pointille]{}}}
        child {node[un] {} {child[pointille]{}}}}
      child {node[zero] {}
        child {node[un] {} {child[pointille]{}}}
        child {node[un] {} {child[pointille]{}}}}}}
  child {node[un] {}
    child {node[un] {}
      child {node[un] {}
        child {node[un] {} {child[pointille]{}}}
        child {node[un] {} {child[pointille]{}}}}
      child {node[un] {} 
        child {node[un] {} {child[pointille]{}}}
        child {node[un] {} {child[pointille]{}}}}}
    child {node[un] {} 
      child {node[un] {} 
        child {node[un] {} {child[pointille]{}}}
        child {node[un] {} {child[pointille]{}}}}
      child {node[un] {}
        child {node[un] {} {child[pointille]{}}}
        child {node[un] {} {child[pointille]{}}}}}};
\end{tikzpicture}

    \end{tabular}
    &
    \begin{tabular}{c}
      \begin{tikzpicture}[->,auto,node distance=2cm,shorten >=1pt,
  semithick,yscale=0.8]
  
  \node[zeroGraph, minimum size=3mm] (A) {1};
  \node (FIRST) [left of=A, node distance=0.7cm] {};
  \node[unGraph,minimum size=3mm] (B)  at (3,-1) {4};
  \node[zeroGraph,minimum size=3mm] (C) at (1.5,-1) {3};
  \node[unGraph,minimum size=3mm] (D) at (1.5,1) {2};
  
  \path (FIRST) edge (A);

  \path (A) edge (D)
  (A) edge (C)
  (B) edge[bend left] (C)
  (B) edge[in=-40,out=-140] (C)
  (C) edge[in=140,out=40] (B)
  (C) edge[bend left] (B)
  (D) edge[loop right,->] ()
  (D) edge[loop above,->] ();

\end{tikzpicture}

    \end{tabular}
  \end{tabular}
  \caption{A rational tree made of two distinct subtrees and its associated multi-graph}
  \label{fig:rat}
\end{figure}

It is possible to characterize rational trees using their complexity.

\begin{theorem}
  \label{th:rational}
  The following proposition are equivalent
  \begin{enumerate}
  \item the tree $T$ is rational;
\item there exists $n$ such that ${\cal P}(n) \leq n $;
\item there exists $n$ such that ${\cal P}(n)  = {\cal P}(n+1)$;
\item There exists $B$ such that for all $n$, ${\cal P}(n) \leq B$.
  \end{enumerate}
\end{theorem}

\begin{proof}
  The proof of this results is similar to the proof for words.\\
1 implies 2:
If $G(T)$ is finite, then the number of factors of size $n$ in $T$ is
smaller than the size of $G(T)$, therefore, there exists $n$ such that
${\cal P}(n) \leq n$.\\
2 implies 3: 
Since ${\cal P}(1) = 2$ and ${\cal P}(n) \leq n$ and since ${\cal  P}$ is non-decreasing with
$n$, there  exists $1 < k <  n$ such that ${\cal P}(k) = {\cal P}(k+1)$.\\
3 implies 4:
If ${\cal P}(n) ={\cal  P}(n+1) = p$ then 
let us call by $A_1^n,\ldots A_p^n$ all the distinct factors of size
$n$ in $T$.
Since $P(n+1) = p$, each $A_i^n$ is prolonged  in a unique way into a
tree of size $n+1$, called $A^{n+1}_i$. 
Now, each sub-tree $A^{n+1}_i$ is composed of a root and $d$ factors
  of  size $n$, in the set  $\{ A_1^n,\ldots A_p^n \}$.  
In turn, they are all prolonged into trees of size $n$ in a unique 
way. Therefore, ${\cal P}(n+2) = p$.
By a direct induction, ${\cal P}(k) = p$ for all $k \geq n$.\\
4 implies 1:
If the number of factors of size $n$ is smaller than $B$ for all $n$,
then this means that the number of equivalence classes for $\equiv_n$
is smaller than $n$ for all $n$, this means that $G(T)$ has less than
$B$ nodes.
\end{proof}

\subsection{Density of rational trees}

Let $T$ be a rational tree and let $G(T)$ be its minimal multi-graph.
The nodes of $G(T)$ are numbered $v_1\cdots,v_K$, with $v_1$ corresponding
to the root of $T$.

$G(T)$ can be seen as the transition kernel of a Markov chain
by considering each arc of $G(T)$ as a transition with probability
$1/d$.

If $G(T)$ is irreducible then the Markov chain  admits a unique stationary measure 
$\pi$ on its nodes.
The density of $T$ and the stationary measure $\pi$ are related by
the following theorem.

\begin{theorem}
  Let $T$ be an irreducible  rational tree with a minimal multigraph
$G(T)$ with $K$ nodes. Let $\ell = (\ell_1, \ldots \ell_K)$ be
  the labels of the nodes of $G(T)$ and let $\pi = (\pi_1,\ldots ,\pi_K)$ be the
  stationary measure over the nodes of $G(T)$. \\
If $T$ is aperiodic, then $T$ admits a density $\alpha = \pi \ell^t$.\\
If $T$ is periodic with period $p$ then $T$ admits an average density
$\alpha = \pi \ell^t$.
\end{theorem}

\begin{proof}
  Let $V_n$ be a Markov chain corresponding to $G(T)$.
Since $G(T)$ is irreducible,  $V_n$ admits a unique stationary measure ,
say $\pi = (\pi_1,\ldots ,\pi_K)$.
Let us call $P$ the kernel of this Markov chain: $P_{i,j} = a/d$ if 
there are $a$ arcs in $G(T)$ from $v_i$ to $v_j$.

Now, let us consider all the  paths  of length $n$ in $T$, starting
from an arbitrary  node $v_i$.
By construction of $G(T)$, the number of  paths that end  up in
the nodes  $v_1,\cdots,v_K$  respectively, of $G(T)$,  is given by the
vector $e_i d^n P^n$, where $e_i$ is the vector with all its coordinates
equal to 0 except the $i$th coordinate, equal to 1.

Now, the number of ones in the tree of height $n$ starting in $v_i$
is $h_n(v_i) = e_i  \sum_{k=0}^{n-1}  d^k P^k \ell^t$.

Let us first consider the case where $P$ is aperiodic.
We denote by $\Pi$ the matrix with all its lines equal to the stationary
measure, $\pi$ and by $D_k $ the matrix $P^k -\Pi$.
When $P$ is aperiodic, then $\lim_{k\to \infty } || D_k||_1 = 0$.
Therefore, for all $k >n,  || D_k||_1 < \epsilon_n  \to 0$.

Then the density of ones  $d_{2n}(v_i) = \frac{d-1}{d^{2n} -1} h_{2n}(v_i)$
can be estimated by splitting the factors of size $2n$ into a factor of size $n$
at the root and $d^^n$ factors of size $n$. One gets 
\begin{eqnarray*} d_{2n}(v_i) & = &   \frac{d-1}{d^{2n} -1} e_i   \sum_{k=1}^n
  d^k P^k \ell^t +  \frac{d-1}{d^{2n} -1} e_i \sum_{k=n+1}^{2n-1} d^k P^k \ell^t,\\
& =&  \frac{d-1}{d^{2n} -1} e_i   (\sum_{k=1}^n d^k P^k + \sum_{k=n+1}^{2n-1}
d^k D_k  + \sum_{k=n+1}^{2n} d^k \Pi)  \ell^t.
\end{eqnarray*}

when $n$ goes to infinity, the first term 
goes to $0$ because $e_i \sum_{k=1}^nd^k P^k \ell^t  \leq d^{n+1}$.
As for the  second term $ \frac{d-1}{d^{2n} -1} e_i \sum_{k=n+1}^{2n-1} d^k D_k
\ell^t \leq  \frac{1}{d^{2n} -1}  d^{2n} \epsilon_n $. This goes to 0 when $n$
goes to infinity.

As for the last term,
$ \frac{d-1}{d^{2n} -1} e_i \sum_{k=n+1}^{2n-1} d^k \Pi   \ell^t
=  \frac{1}{d^{2n} -1}  ( d^{2n} - d^{n+2}) (e_i \Pi) \ell^t$ this goes
to $\pi\ell^t$ when $n$ goes to infinity.

The same holds by computing the density of trees of size $2n+1$
by splitting them into the first $n+1$ levels and the last $n$ levels.

This shows that the rooted density of all the trees in $T$ is the same,
equal to  $\pi\ell^t$.
\medskip

Let us now consider the case when the tree is periodic with period $p$.
In that case, the kernel of $p$ steps of the Markov chain can be put
under the form
\[ P^p = 
 \left[
\begin{array}{c|c|c|c}
P_1 & 0 & \cdots & 0  \\ \hline
0 & P_2 & \ddots  & 0   \\ \hline
\vdots & \ddots  & \ddots & \vdots   \\ \hline 
0 & 0 &  \ldots & P_m
\end{array}
\right]. \]

The sub-matrices $P_1, \ldots , P_m$ are the kernels of aperiodic chains defined on a
partition $S_1\ldots S_m$ of the nodes of $G(T)$.
Let us denote  by $\alpha_1, \ldots \alpha_m$ the densities of the factors of
size $np$, starting in $S_1\ldots S_m$, respectively (they exist because
this has just been proved  for aperiodic trees).

Starting from a node $v$  the average density of a tree of
size $n = pq_n +r_n$, $r_n < p$ is  
\[ \frac{1}{n} \sum_{k=0}^n d_k(v) =  \frac{1}{pq_n+r_n} (\sum_{a=0}^{q_n}\sum_{b=0}^p
d_{ap+b+r}(v) +  \frac{1}{pq_n+r_n} \sum_{k=0}^{r_n}  d_k(v) ).\]

The  first term goes to  $(\alpha_1+ \ldots+\alpha_m)/m$ while the second term goes
to zero, when $n$ goes to infinity, independently of the root.
Finally, $(\alpha_1+ \ldots+\alpha_m)/m = (\pi_1' \ell_1^t+ \cdots + \pi_m'\ell_m^t)/m  = \pi \ell^t$
where $\pi_1' , \cdots , \pi_m'$ are the stationary probability for the kernels
$P_1, \ldots , P_m$ and $\ell_1, \ldots \ell_m$ are  the vectors of the  labels in 
 $S_1\ldots S_m$.



\end{proof}

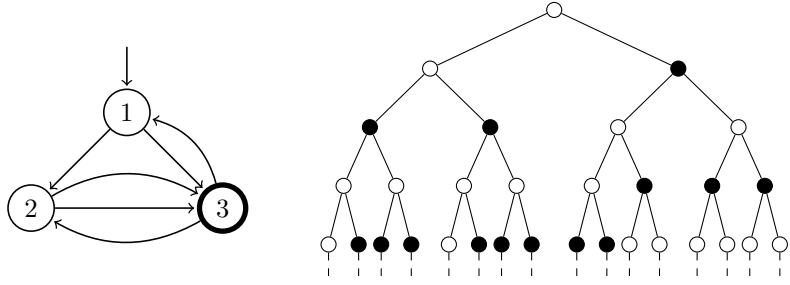
\begin{figure}[ht]
  \centering
  \begin{tabular}{cc}
    \begin{tabular}{c}
      \begin{tikzpicture}[->,auto,node distance=1.8cm,shorten >=1pt,
  semithick]
  
  \node[zeroGraph,minimum size=3mm] (A) {1};
  \node (FIRST) [above of=A, node distance=1.cm] {};
  \node[zeroGraph,minimum size=3mm] (B) [below left of=A] {2};
  \node[unGraph,minimum size=3mm] (C) [below right of=A] {3};
  
  \path (FIRST) edge (A);
  
  \path (A) edge     (B)
  (A) edge (C)
  (B) edge             (C)
  (B) edge [bend left] (C)
  (C) edge [bend right] (A)
  (C) edge [bend left] (B);
\end{tikzpicture}
    \end{tabular}&
    \begin{tabular}{c}
      \begin{tikzpicture}[yscale=0.6]
  
  \node[zero] {}
  child {node[zero] {} 
    child {node[un] {} 
      child {node[zero] {} 
        child {node[zero] {} {child[pointille]{}}}
        child {node[un] {} {child[pointille]{}}}}
      child {node[zero] {} 
        child {node[un] {} {child[pointille]{}}}
        child {node[un] {} {child[pointille]{}}}}}
    child {node[un] {} 
      child {node[zero] {} 
        child {node[zero] {} child[pointille]{}}
        child {node[un] {} {child[pointille]{}}}}
      child {node[zero] {}
        child {node[un] {} {child[pointille]{}}}
        child {node[un] {} {child[pointille]{}}}}}}
  child {node[un] {}
    child {node[zero] {}
      child {node[zero] {}
        child {node[un] {} {child[pointille]{}}}
        child {node[un] {} {child[pointille]{}}}}
      child {node[un] {} 
        child {node[zero] {} {child[pointille]{}}}
        child {node[zero] {} {child[pointille]{}}}}}
    child {node[zero] {} 
      child {node[un] {} 
        child {node[zero] {} {child[pointille]{}}}
        child {node[zero] {} {child[pointille]{}}}}
      child {node[un] {}
        child {node[zero] {} {child[pointille]{}}}
        child {node[zero] {} {child[pointille]{}}}}}};
\end{tikzpicture}
    \end{tabular}
  \end{tabular}
  \caption{An irreducible aperiodic  rational tree and its 
    minimal graph. The stationary probabilities over the associated
    Markov  chain are $\pi = (2/9,3/9,4/9)$.  The density of the tree
  is $\alpha = 4/9$.}
  \label{fig:density_example}
\end{figure}

An example illustrating the computation of the density of an aperiodic
irreducible rational tree is given in Figure \ref{fig:density_example}.
The stationary measure of the Markov chain is $\pi = (2/9,3/9,4/9)$.
Therefore, the density is $\alpha = 2/9\ell_1 + 3/9\ell_2 + 4/9\ell_3 =
4/9$.

As for the reducible case,
it should be easy to see that a rational tree may have different
(average) densities for some of its subtrees  (this is the case for
the leftmost 
tree   in Figure \ref{fig:having_a_density}).
Therefore, a reducible tree does not have a density nor an average
density in general.

Let us call $S_1,\cdots,S_m$ the final strongly connected components of
$G(T)$. Let $\alpha_1,\ldots,\alpha_m$ be the average densities of the
components $S_1\ldots,S_m$ respectively.
Finally, let $R = (R_1\cdots R_m)$ be the probability of reaching the
components $S_1\cdots S_m$ starting from
the root $v_1$, in the Markov chain associated with $G(T)$.  
Then, the following is true.  

\begin{theorem}\label{th:reduc}
  A rational tree always has a rooted average density $\alpha = 
(\alpha_1,\ldots \alpha_m) R^t$.
\end{theorem} 

\begin{proof} 
If  $P$ is reducible, $P$ can be decomposed into
\begin{equation*}
P = \left[
\begin{array}{c|c|c|c}
Q& K_1& \cdots & K_m   \\ \hline
0 & P_1 & \cdots & 0   \\ \hline
\vdots & \ldots  & \ddots & \vdots   \\ \hline 
0 & 0 & \ldots & P_m
\end{array}
\right]
\mbox{ and }
P^n = \left[
\begin{array}{c|c|c|c}
Q^n& K_1'& \cdots & K_m' \\\hline
0 & P_1^n & \cdots & 0 \\\hline
\vdots & \ldots  & \ddots & \vdots \\ \hline
0 & 0 & \ldots & P_m^n
\end{array}
\right]
\end{equation*}

Considering all the paths in $G(T)$ of length $n$, starting in the root,
the number of paths ending in component $S_\ell $ is
$N_\ell(n)  = d^n \sum_{i\in S_\ell} P^n_{1i}$.
Let us decompose all  the paths ending  in $S_\ell$ into two sub-paths:
one (of length $k$) before entering $S_\ell$ and one (of length $n-k$)
inside $S_\ell$, we get from the decomposition of $P^n$, 
$N_\ell (n) =d^n \sum_{k=0}^n (1,0,\ldots,0) Q^k K_\ell u_\ell $, where $u_\ell $ is a
vector whose coordinates are 1 in $ S_\ell $ and 0 everywhere else.

The number of 1 in the rooted subtree of $T$ of size $2n$
is the number of ones in all the paths of length $n$ plus the number
of ones in the subtrees of size $1$.
When $n$ is large, the number of ones in the paths can be neglected
with respect to the number of ones in the end trees.

Finally, the number of one in a tree of size $2n$ is the number of ones
in each possible  end-tree of size $n$  times the number of such
trees, namely  $N_\ell (n)$.
When $n$ goes to infinity,  the density of ones goes to  
$\sum_{\ell=1..m}  \alpha_\ell   (1,0,\ldots,0) (I-Q)^{-1} K_\ell u_\ell = (\alpha_1 \cdots \alpha_m) R^t$,
with $R_\ell  =  (1,0,\ldots,0) (I-Q)^{-1}  K_\ell u_\ell$.
\end{proof} 

An example illustrating the computation of the rooted average density
of a tree is given in Figure \ref{fig:rat}.
The graph $G(T)$ has two final components, one aperiodic component with  density 1 and
another one with period 2 with average density 1/2.
Starting from the root, both components are reached with probability
1/2.
Therefore, such a tree has an average rooted density
$\alpha = 1/2 (1/2) + 1/2 (1) = 2/3$.

Also, it is not difficult to show that if all final component have a
density (rather than an average density), then the tree has a rooted
density, given by the same formula as in Theorem \ref{th:reduc}.

Finally, it is fairly straightforward  to prove
 that since the  $K \cdot K$ kernel $P$ of the Markov
chain associated with $G(T)$ has all its elements of the form $a/d$,
then the stationary probabilities $\pi$ as well as the average rooted density $\alpha$ of a
rational tree are rational numbers of the form $c/b$ with $ 0\leq c
\leq b \leq d^{K+1}$.
This fact will be used in the algorithmic section \ref{sec:algo} to
make sure that the complexity of the algorithms does not depend on the
size of the numbers.

\section{Balanced and Mechanical Trees}



In this section, we will introduce our most important definitions: strongly
balanced and mechanical trees and explore the relations between them.  In
particular we will prove that in the case of irrational trees they
represent the same set of trees, giving us a constructive representation of
this class of trees.  These results are very similar to the ones on words,
which are summarized below.

\subsection{Sturmian, Balanced and Mechanical Words\label{sec:words}}

One  definition of a Sturmian word uses  the complexity of a
word.  The complexity of an infinite word $w$ is a function
${\cal P}_w:\mathbb{N}\to\mathbb{N}$ where ${\cal P}_w(n)$ is the number of distinct
factors of length $n$ of the word $w$. A word is periodic if there exists
$n$ such that ${\cal P}_w(n)\leq n$.  Sturmian words are aperiodic words with
minimal complexity, \emph{i.e} such that for any $n$: 
\begin{equation}
{\cal P}_w(n) = n+1. \label{eq:sturmian_word}
\end{equation}
If $x$ is a factor of $w$, its height $h(x)$ is the number of letters equal
to $1$ in $x$. A balanced word is a word where the letters $1$ are
distributed as evenly as possible:
\begin{equation}
  \label{eq:balanced_word}
 \forall  x,y\mathrm{~factors~of~} w, |x|=|y| \Rightarrow |h(x)-h(y)|\leq 1.
\end{equation}
A mechanical word is constructed using integer parts of affine
functions.  Let $\alpha\in[0;1]$ and $\phi\in[0;1)$. The
lower (resp. upper)  mechanical word of {\it slope} $\alpha$ and {\it
  phase}  $\phi$,  $w=w_1w_2\dots$  (resp. $w'=w_1'w_2'\dots$) is defined
by:  
\begin{eqnarray}
  \label{eq:mechanical_word}
  \forall i\geq 1
  \left. \begin{array}{l}
    w_i = \lfloor (i+1)\alpha+\phi\rfloor - \lfloor
    n\alpha+\phi\rfloor,\\
    w'_i = \lceil (i+1)\alpha+\phi\rceil - \lceil i\alpha+\phi\rceil.
  \end{array} \right.
\end{eqnarray}

These  three definitions represent almost the same set of words. In the case of
aperiodic words, they are equivalent: a word is Sturmian if and only
if it is
balanced and aperiodic if and only if  it is mechanical of irrational slope. For periodic words, there are similar
relations:
\begin{itemize}
\item A rational mechanical word is balanced.
\item A periodic balanced word is ultimately mechanical.
\end{itemize}
A word is an ultimately mechanical word if it can written as $xw$ where $x$
is a finite word and $w$ is a mechanical word. An example of a balanced 
word which is not mechanical (and just ultimately mechanical) is the
infinite word with all letter $0$ and just one letter $1$. For a more
complete description of Sturmian words, we refer to
\cite{lothaire2002acw}. 

\subsection{Balanced and strongly balanced trees}

Using the two definitions of factors of a tree, we define two notions of
balance for trees: the first one and probably the most natural one, is what
we call \emph{balanced trees} and the other one is called \emph{strongly balanced
  trees}. 

\begin{defi}[Balanced and strongly balanced tree]
  A tree is balanced if for all $n\geq 0$, the number of nodes label by $1$
  in two factors of size $n$ differ by at most $1$.
  
  A tree is strongly balanced if for all $n,k>0$, the number of $1$ in two
  factors  of size $n$ and width $k$ differ by at most $1$. 
\end{defi}

As the name suggests, the strong  balance property implies  balance  (by taking $k=1$). In fact this notion is strictly
stronger, see section \ref{sec:glossary} for an example of balanced tree
that are not strongly balanced. 

Although the balance property  is weaker and seems more natural for a
generalization from words, our results will be mainly focused on strongly
balanced trees that have almost the same properties as their  counterparts on
words. 

\subsubsection{Density of a balanced tree}

For all node $v$ and all size $n$, we denote by  $h_v(n)$ the number of $1$
in the factor of root $v$ of size $n$ and $d_v(n)$ the density of
this subtree is the number of ones divided by the cardinal $S(n)$ of the
factor:  $d_v(n) \bydef \frac{1}{\Sn{n}}h_v(n)$. 

\begin{prop}[Density of balanced tree]
  A balanced tree has a density $\alpha$.
  
  Moreover for all node $v$ and for all size $n$:
  \begin{equation}
    |h_v(n) - \lfloor \Sn{n}\alpha\rfloor |\leq 1\label{eq:densite_balance} 
  \end{equation}
\end{prop}


\begin{proof} Let $m_n$ be the minimal number of $1$ in all factors of
  size $n$.  As the tree is balanced, for all nodes $v$ and $n\geq 1$:
  \begin{equation}
    m_n \leq h_v(n) \leq m_n+1
  \end{equation}
  Now let us consider a factor of size $n+k$ and root $v$. It can be
  decomposed in a factor of size $k$ of root $v$ and $d^k$ factors
  of size $n$ at the leaves of the previous factor. The number of ones in
  these factors can be bounded by $m_n$ and $m_k$, therefore we have:
  \begin{equation}
    m_k + d^k m_n \leq m_{n+k}\leq m_k+1+d^k(m_n+1) \label{eq:bal_dens1} 
  \end{equation}
  
  The density of a factor of size $n$ is 
  $\frac{m_n}{\Sn{n}} \leq d_v(n) = \frac{h_v(n)}{\Sn{n}} \leq
  \frac{m_n+1}{\Sn{n}}$. Using these facts, we can bound
  $d_v(n+k)-d_v(n)$: 
  \begin{equation*}
    \frac{m_{n+k}}{\Sn{n+k}} - \frac{m_n+1}{\Sn{n}}
    \leq d_v(n+k)-d_v(n) \leq \frac{m_{n+k}+1}{\Sn{n+k}} -
    \frac{m_n}{\Sn{n}}
  \end{equation*}
  Using \eqref{eq:bal_dens1}, the left inequality can be lower bounded by
  \begin{eqnarray*}
    (d-1)\big( \frac{d^km_n+m_k}{d^{n+k}-1} - \frac{m_n+1}{d^{n}-1}\big)
    &=&(d-1)\big( \frac{m_n+m_k/d^k}{d^n-1/d^k} -
    \frac{m_n+1}{d^{n}-1}\big)\\
    &\geq&(d-1)\big( \frac{m_n}{d^n-1}-\frac{m_n+1}{d^{n}-1}\big)\\
    &\geq&-\frac{1}{\Sn{n}}
  \end{eqnarray*}
  
  The same method can be used to prove  that $d_v(n+k)-d_v(n)\leq
  \frac{1}{S(n)}$, which shows that for $n$ big enough,
  $|d_v(n+k)-d_v(n)|$ is smaller than $\epsilon$, regardless of $k$. 
  Thus $d_v(n)$ is a Cauchy sequence and has a limit
  $\alpha=\lim_{n\to\infty}\frac{m_n}{S(n)}$. This limit does not depend
  on $v$ and the tree has a density.
  
  Lets now prove that $d_v(n) - \lfloor \Sn{n}\alpha  \rfloor | \leq 1$:
  dividing the inequality \eqref{eq:bal_dens1} by $\Sn{n,k}$ and taking 
  the limit when $k$ tends to $\infty$ leads to: 
  \begin{equation*}
    \frac{(d-1)m_n+\alpha}{d^n} \leq \alpha \leq  \frac{(d-1)m_n+1+\alpha}{d^n}.
  \end{equation*}
  
 This  shows that: $\Sn{n}\alpha -1 \leq m_n \leq
  \Sn{n}\alpha$, which implies Equation \eqref{eq:densite_balance}.  
  
\end{proof}

Similar ideas can be used  to show that 
Equation \eqref{eq:densite_balance} can be improved in the case of
strongly balanced tree: for all width and size  $k,n\geq 1$, the number of ones
$h(n,k)$ in a factor of size $n$ and width $k$ satisfies:
\begin{eqnarray}
  \big|h(n,k)-\lfloor\Sn{n,k}\alpha\rfloor \big|\leq 1\label{eq:strong_bal} 
\end{eqnarray}

\subsection{Mechanical trees}

Building balanced tree is not that easy. According to formula
\eqref{eq:densite_balance}, each factors of size $n$ must have $\lfloor
\alpha\Sn{n}\rfloor$ or $\lfloor\alpha\Sn{n}\rfloor+1$ nodes one. This
leads us to the following construction, inspired by the construction of
mechanical words.

\begin{defi}[Mechanical tree]
  \label{def:mechanical_tree}
  A tree is mechanical of density $\alpha\in[0;1]$ if for all node $v$,
  there exists a phase $\phi_v$ which satisfies one
  of the two following properties:
    
  \begin{eqnarray}
    \forall n: h_v(n) = \bigg\lfloor\Sn{n}\alpha+\phi_v
    \bigg\rfloor,\label{eq:mecha_inferior}\\
    \mathrm{or~}\forall n: h_v(n) = \bigg\lceil\Sn{n}\alpha-\phi_v
    \bigg\rceil. \label{eq:mecha_superior} 
  \end{eqnarray}
  
  In the first case, we say that $\phi_v$ is \emph{an} inferior phase of
  $v$. In the second case, we say that $\phi_v$ is \emph{a} superior phase
  of $v$. 
\end{defi}

This definition suggests that the phases of all nodes could be
arbitrary. In fact, we will see that there exists  a unique mechanical tree
with a given phase at the root.  The second question raised by this definition
is the existence and uniqueness of the phase: we call $\phi_v$ ``a'' phase of a node
$\phi_v$ and not ``the'' phase of $\phi_v$ since
there may exist several phases leading to the same tree. 
\medskip

We begin by a characterization of mechanical trees, given in  the following
formula: 
\begin{prop}[Characterization of mechanical trees]
  For each $\alpha\in[0;1]$ and $\phi\in[0;1)$, there exists a unique
  mechanical tree of density $\alpha$ such that $\phi$ is an inferior
  (resp. superior) phase of the root.
  
  Moreover, if $\phi$ is an inferior (resp. superior) phase of a node then
  $\phi_0\leq\dots\leq\phi_{d-1}$ are inferior (resp. superior) phases of
  its $d$ children, with
  \begin{equation}
    \phi_i = \frac{\alpha+\phi+i-\lfloor \alpha+\phi\rfloor}{d}
    \quad \left(\mathrm{resp.~} \phi_v=\frac{\phi+\lceil \alpha-\phi\rceil - \alpha
      +i}{d}\right).
    \label{eq:relation_between_phases}
  \end{equation} 
  \label{prop:uniq}
\end{prop}

The proof will be done in two steps. First we will see that if we define
the phases as in \eqref{eq:relation_between_phases} we have a mechanical
tree, then we will see that this is the only way to do so. 

\begin{proof}
  \textbf{Existence. }
  Let $\alpha\in[0;1]$ and $\phi\in[0;1)$. We want to build a mechanical
  tree which root has an inferior phase $\phi$ ( the case of a superior
  phase if similar and is not detailed here). Let $\mathcal{A}$ be an infinite tree. To each node
  $v$, we associate a number $\phi_v$ defined by:
  \begin{itemize}
  \item $\phi_\mathrm{root}=\phi$.
  \item If the phase of a node $v$ is $\phi_v$, its $d$ children satisfy
    Equation \eqref{eq:relation_between_phases}.
  \end{itemize}
  Then we build a labeled tree by associating to each node $v$ the label
  $\lfloor \alpha+\phi_v\rfloor$. Let us prove by induction on $n$ that the following
  relation holds.
  \begin{equation}
    \mathrm{For~all~v:~} h_v(n) =
    \bigg\lfloor\Sn{n}\alpha+\phi_v\bigg\rfloor.\label{eq:induction} 
  \end{equation} 
  By definition of the labels, \eqref{eq:induction} holds when $n=1$. Let
  $n\geq 0$ and let us assume that \eqref{eq:induction} holds for $n$. Let $v$ be a node with phase $\phi_v$ and let
  $\phi_0\dots\phi_{d-1}$ be the phases of its children. We assume that
  $\alpha+\phi_v<1$ (which means that the label of the node is $0$)
  a similar calculation can be done in the other case,  $\alpha+\phi_v>1$.

  Using the well-known formula $\sum_{i=0}^{d-1} \lfloor x+\frac{i}{d}
  \rfloor = \lfloor dx \rfloor$, we can compute $h_v(n+1)$:  
  \begin{eqnarray*}
    h_v(n+1) &=& \sum_{i=0}^{d-1}
    \lfloor \Sn{n}\alpha+\phi_i\rfloor\\ 
    &=& 
    \sum_{i=0}^{d-1} \lfloor
    \frac{d^{n}-1}{d-1}\alpha+\frac{\alpha+\phi+i}{d} \rfloor
    \label{eq:mecha_1}\\
    &=&\lfloor d(\frac{d^{n}-1}{d-1}\alpha+\frac{\alpha+\phi}{d})\rfloor
    \label{eq:mecha_2}\\
    &=& \lfloor  \Sn{n+1}\alpha+\phi \rfloor.
  \end{eqnarray*}
  Therefore, \eqref{eq:induction} holds for all $n$ which means that the
  tree is mechanical. 
  
  \textbf{Uniqueness. } Now, let $\mathcal{A}$ be a mechanical tree of
  density $\alpha$. Let $v$ be a node and $\phi_0,\dots,\phi_{d-1}$ be the phases
  of its children. Let $i$ and $j$ be two children and let $h_i(n)$ be the
  number of ones  in the  $i$th child  subtree (of phase $\phi_i$). We want to prove
  that either for all n: $h_i(n)\leq h_j(n)$ or for all $n$: $h_i(n) \geq
  h_j(n)$.  If the two nodes are both inferior (resp. superior), this is
  clearly true: $h_i(n)\leq h_j(n)$ if and only if $\phi_i \leq \phi_j$
  (resp. $\phi_i\geq\phi_j$). If $i$ is inferior and $j$ is superior, it is not
  difficult to 
  show that $\phi_i<1-\phi_j$ implies $h_i(n)\leq h_j(n)$ and
  $\phi_i\geq 1-\phi_j$ implies $h_i(n)\geq h_j(n)$.

  Therefore we can assume (otherwise we exchange the order of the
  children) that for all $n$:
  \[ h_0(n)\leq h_1(n)\leq \dots \leq h_{d-1}(n).\]
  Moreover  as $h_{d-1}(n)-h_0(n) \leq 1$, there exists $k$ such that $h_0(n) =
  h_1(n)=\dots=h_k(n)<h_{k+1}(n) = \dots = h_{d-1}(n)$. As
  $\sum_{i=0}^{d-1}h_i(n)$ does not depend on $\phi_0,\ldots,\phi_{d-1}$, then for each $n$ there 
  is only one $k$ that works and therefore there are only one possibility
  for $h_i(n)$ for all $n$ and all $i$. This implies that the tree with root $v$ is unique
  
  As we have seen in the beginning of the proof, the phase  $\phi_i$ defined in
  \eqref{eq:relation_between_phases} defines  correct values for  $h_i(\cdot)$. Therefore
  such a phase $\phi_i$ is  a possible phase  for  the $i$th child.
\end{proof}

  

This theorem shows that when the phase is fixed the tree is unique. The
converse  is false and one can find  several phases that lead to the same
tree (for example,  when $\alpha=0$ all phases define the tree with label
0 everywhere) but we will show next   that the set of
densities $\alpha$ for which the phases  are  not necessarily unique has 
Lebesgue measure zero.

If for all $n$, $\Sn{n}\alpha+\phi\not\in \N$, then $\lfloor
\Sn{n}\alpha+\phi \rfloor = \lceil \Sn{n}\alpha+\phi-1\rceil$. In that case,
if $\phi$ is an inferior phase of a node then $1-\phi$ is a superior phase
of the node. Therefore except particular cases, there exists at least two
phases of a node: one inferior and one superior. Let us now look at the
possible uniqueness of the inferior phase.

Let us call $\mathrm{frac}(x)$ the fractional part of a real number $x$ and
let us consider  the sequence
$\left\{\mathrm{frac}(\Sn{n}\alpha+\phi)\right\}_{n\in \N}$. If this sequence can be
arbitrary close to $0$, this means that for all $\psi<\phi$, there exists $k$
such that $\lfloor\Sn{k}\alpha+\psi\rfloor<\lfloor\Sn{k}\alpha+\phi\rfloor$
and $\psi$ can not be a phase of the tree. Also,  if this sequence can be
arbitrary close to $1$, then one can show similarly that for all $\psi>\phi$, $\psi$
is not a phase of the node. 
Conversely, if the exists $\delta >0$ such that
$ \mathrm{frac}(\Sn{n}\alpha+\phi) > \delta$ (resp. $< 1-\delta$) for all $n$, then
let $\phi' = \phi - \varepsilon$ (resp.  $\phi' = \phi + \varepsilon$), with $\varepsilon  < \delta$.
Then  $\lfloor\Sn{n}\alpha+\phi \rfloor = \lfloor\Sn{n}\alpha+\phi' \rfloor$ for all $n$.

Finally, a phase $\phi$ is unique  if and only if 
$0$ and $1$ are accumulation points  of the sequence $\left\{\mathrm{frac}(\Sn{n}\alpha+\phi)\right\}_{n\in \N}$.

Let us call $x\bydef \frac{1}{d-1}\alpha$ and $y \bydef  \phi-x$ and let us consider the
sequence 
\[\mathrm{frac}(\Sn{n}\alpha+\phi)=\mathrm{frac}(xd^n - y). \]
Let $x_1,\dots,x_k,\dots$ (resp. $y_1,y_2,\dots$) be the sequence of
the  digits of $x$ (resp. $y$) in base $d$ (also called the $d$-decomposition). We have:
\[ xd^n - y =
\underbrace{\sum_{k=1}^nx_kd^{n-k}}_{\in \mathbb{N}}+\sum_{k=1}^\infty
(x_{k+n}-y_k)d^{-k} \]
$\mathrm{frac}(xd^n-y)$ is arbitrarily close to $0$ implies that for arbitrarily big $k$,
there exists $n$ such that 
\begin{equation}
\label{p1}x_{n},\dots,x_{n+k-2} = y_1,\dots,y_k, \quad x_{n+k-1}>y_n,
\end{equation}
or 
\begin{equation}
\label{p2}
 \mathrm{frac }(xd^n - y) = 0.
\end{equation}
Also, $xd^n-y$
is arbitrarily close to $1$ implies that for arbitrarily big $k$, 
there exists $n$ such that 
\begin{equation}
\label{p3}
x_{n},\dots,x_{n+k-2} = y_1,\dots,y_{n-1} , \quad  x_{n+k-1} < y_n,
\end{equation}
or the $d$-development of $y$ is finite (with only
zeros after some point $\ell: y =  y_1,\dots,y_\ell , 1, 0, 0\ldots $) and  that for arbitrarily big $k$,  
 there exists $n$ such that 
\begin{equation}
\label{p4} x_{n},\dots,x_{n+k-2} =
 y_1,\dots,y_{\ell},0, 1,\ldots, 1.
\end{equation}

Using this characterization, three cases can be distinguished.

\begin{itemize}
\item If $\frac{\alpha}{d-1}$ is a number such that all finite sequences
  over $0,\ldots,d-1$ 
  appear  in its $d$-decomposition, then every phase is
  unique. In particular, all \emph{normal numbers}\footnote{A number
    is normal in base $d$ if all sequences of
  length $k$ appear uniformly in its
  $d$-decomposition}
 in base $d$ verify this property and
  it is known that almost every number in $[0,1]$ is normal (see
  \cite{borel1909ldl} or \cite{durrett1991pta}).
\item If $\alpha \in \mathbb{Q}$, then the sequence
  $\mathrm{frac}(\Sn{k}\alpha+\phi)\big)$ is periodic and there are no
  phase $\phi$ such that $\phi$ is unique. 
\item If $\alpha$ is neither rational nor has the property that all
  binary sequences  appear in $\alpha$, then some $\phi$ can be
unique and some others may not. For example, for $d=2$, if $\alpha$ is
  (in base 2) the number  
  \[ \alpha = 0.101100111000111100001111100000\dots,\]
then if $\mathrm{frac}(\alpha-\phi)=0$, $\phi$ is unique (because $\alpha$ satisfies
both properties  \ref{p1} and \ref{p4}). However  $\phi_1$
  and $\phi_2$ such that $\mathrm{frac}(\alpha-\phi_1)=0.10100$ and
  $\mathrm{frac}(\alpha-\phi_2)=0.1010$ are equivalent  (generate the same tree).

  Other examples of the same type are 
  the \emph{rewind trees}, drawn on figure \ref{fig:irrat_mecha}. The
  sequence of digits in base $2$ of the density of trees  is a
  Sturmian word  with
  irrational density.  Half of the nodes of the tree are associated
  with node  $0$ in the minimal graph and
  therefore could have the  same phase  whereas the phases computed using
  Equation \ref{eq:relation_between_phases} are not all the same.
  Therefore, phases are not unique here.
\end{itemize}

\subsubsection{Phases of a tree\label{sec:all_phases}}

Let us call $\Phi_v$ the set of numbers that can be phases of a node $v$
and $\Phi$,  the set of the possible phases of a tree is the union of all
possible phases of its nodes: $\Phi=\cup_b\Phi_v$. The set $\Phi$ may be countable
or uncountable. Countable for 
example when $\alpha/(d-1)$ is normal since there are at most as many
phases as nodes. Uncountable for example for the tree with all label $0$,
for which for each node, all phases in $[0;1)$ work. Nevertheless, the set
of possible phases is dense is $[0;1)$.  

Indeed, at least all phases defined by the relation
\eqref{eq:relation_between_phases} are in $\Phi$. If $\phi$ is the phase of
the root, then all nodes at level $k$ have a phase which is the fractional
part of:
\begin{equation}
  \label{eq:all_phases}
  \frac{\frac{\frac{\phi+\alpha+i_k}{d}+\alpha+i_{k-1}}{d}+\dots+\alpha+i_1}{d}
  = \alpha(\frac{1}{d^k}\dots\frac{1}{d}) + \frac{\phi}{d^k} +
  \frac{i_k}{d^k}+\dots+\frac{i_1}{d^1},
\end{equation}
with $0\leq i_j< d$ for all $j$. Conversely all of these numbers are the
phases of some node at level $k$. 

As $k$ tends to infinity, by a proper choice  of $i_1,\dots,i_k$ the
fractional part of this number can be as close as possible to any number in
$[0;1]$.  Thus the set of phases of the tree is dense in $[0;1]$. 

If the density is $\frac{p(d-1)}{d^{n+k}-d^k}$ (with $n+k$ minimal) one can
show that the set of all possible phases for a given node is 
$[\frac{d^m-1}{d-1}\alpha;\min(\frac{d^{m+1}-1}{d-1}\alpha,1))$ for some
$m\in{0,\dots,n+k-1}$. As $\Phi$ is dense in $[0;1)$, it contains all of 
these intervals. Therefore, $\Phi=[0;1)$ and the tree has exactly $n+k$
different factors of size greater than $n+k$. Hence its minimal graph has
exactly $n+k$ nodes.

\subsection{Equivalence between strongly balanced and mechanical trees} 

As we have seen in section \ref{sec:words}, there are strong relations
between balanced and mechanical words. In this part, we will see that we
can prove the same results between strongly balanced and mechanical
trees. This result is formally stated in the following theorem.

A tree is \emph{ultimately mechanical} if
all nodes (except  finitely  many)  are mechanical (\emph{i.e.} satisfies the equations
\ref{eq:mecha_inferior} or \ref{eq:mecha_superior}).

\begin{theorem} The following statements are true.
  \label{th:equiv_mecha_balanc}
  \begin{enumerate}
    \renewcommand{\labelenumi}{\textnormal{(\roman{enumi})}}
  \item A mechanical tree is strongly balanced.
  \item An irrational strongly balanced tree is mechanical.
    \item  A rational strongly balanced tree is ultimately mechanical.
   %
   %
  \end{enumerate}
\end{theorem}

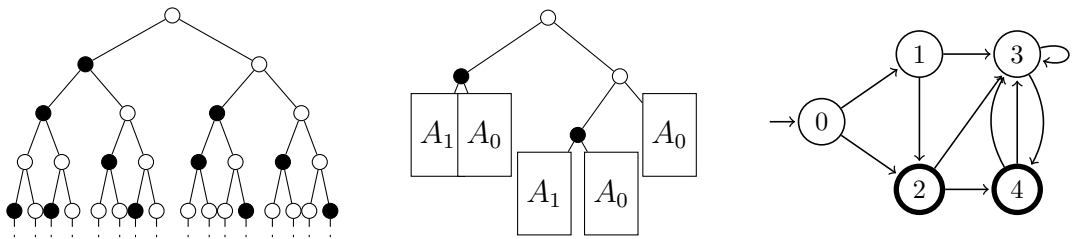
\begin{figure}[ht]
  \begin{center}
    \begin{tabular}{ccc}
      \begin{tabular}{c}
        \begin{tikzpicture}[xscale=0.7, yscale=0.5]
  \node[zero] (root) {} 
  child { node [un]{} 
    child { node [un]{} 
      child { node [zero]{}
        child { node [un]{} child[pointille]{}} 
        child { node [zero]{} child[pointille]{}}} 
      child { node [zero]{}
        child { node [un]{} child[pointille]{}} 
        child { node [zero]{} child[pointille]{}}}}
    child { node [zero]{}
      child { node [un]{}
        child { node [zero]{} child[pointille]{}} 
        child { node [zero]{} child[pointille]{}}} 
      child { node [zero]{}
        child { node [un]{} child[pointille]{}}
        child { node [zero]{} child[pointille]{}}}}} 
  child { node [zero]{} 
    child { node [un]{} 
      child { node [un]{}
        child { node [zero]{} child[pointille]{}} 
        child { node [zero]{} child[pointille]{}}} 
      child { node [zero]{}
        child { node [zero]{} child[pointille]{}} 
        child { node [un]{} child[pointille]{}}}}
    child { node [zero]{} 
      child { node [un]{}
        child { node [zero]{} child[pointille]{}}
        child { node [zero]{} child[pointille]{}}} 
      child { node [zero]{}
        child { node [zero]{} child[pointille]{}} 
        child { node [un]{} child[pointille]{}}}}};
  
    

\end{tikzpicture}
      \end{tabular}
      &
      \begin{tabular}{c}
        \begin{tikzpicture}[yscale=.6, xscale=.7]
  \tikzstyle{noeud} = [circle, draw];
  \tikzstyle{arbre_mechanique} = [draw,rectangle, minimum
  height=1.1cm,fill=white]

  \node[zero] {}
  child {node[un] {}
    child {node[arbre_mechanique,xshift=.25cm]{\large $A_1$}}
    child {node[arbre_mechanique,xshift=-.25cm]{\large $A_0$}}
  }
  child {node[zero, xshift=-.2cm]{}
    child { node[un] {}
      child {node[arbre_mechanique,xshift=-.2cm]{\large $A_1$}}
      child {node[arbre_mechanique,xshift=.2cm]{\large $A_0$}}}
    child {node[arbre_mechanique, xshift=.1cm] {\large $A_0$}}};
     
\end{tikzpicture}
      \end{tabular}
      &
      \begin{tabular}{c}
        \begin{tikzpicture}[auto,node distance=1.3cm,semithick,->,shorten >=1pt,
  scale=0.8]
  
  \node[zeroGraph,minimum size=3mm] (a) {0};
  \node[minimum size=0mm, inner sep=0pt] (x) [left of = a, node
  distance=0.7cm] {};
  \node[zeroGraph,minimum size=3mm] (b) [right of = a, yshift=.9cm] {1};
  \node[unGraph,minimum size=3mm] (c) [right of = a, yshift=-.9cm] {2};
  \node[zeroGraph, minimum size=3mm] (0) [right of = b] {3};
  \node[unGraph,minimum size=3mm] (1) [right of = c] {4};

  \draw (x) edge (a)
  (a) edge (b)
  (a) edge (c)
  (b) edge (0)
  (c) edge (0)
  (c) edge (1)
  (b) edge (c)
  (1) edge[bend left] (0)
  (1) edge (0)
  (0) edge[loop right] ()
  (0) edge[bend left] (1);
  
\end{tikzpicture}
      \end{tabular}
    \end{tabular}
  \end{center} 
  
  \caption{Example of a rational tree that is strongly balanced but not
     mechanical. On the left is the tree itself. In  the middle
     the mechanical suffixes of the tree  are displayed
    and its  corresponding minimal graph (reducible)  is displayed  on the right.
    \newline
    One can verify on the picture that the beginning of this tree 
    is strongly balanced and as it continues with density exactly $1/3$, the
    whole tree is strongly balanced. However this tree is ultimately
    mechanical but not mechanical since in a mechanical tree of density
    $1/3$, all  factors  of size $2$ should have $\lfloor 1+\phi\rfloor = 1$ node
    labeled by one.  }
  \label{fig:ulti_mecha}
\end{figure}

This theorem is the analog of the theorem linking balanced and mechanical
words.  We have seen that the word $0^k10^\infty$ is balanced but not
mechanical, only ultimately mechanical. Its counterpart for trees would be a
tree with all label equal to $0$ except for one node which has a label
$1$.  The number $1$ can be chosen as deep as desired, which shows that we
can not bound the size of the ``non-mechanical'' beginning of the
tree. A more complicated example is drawn Figure \ref{fig:ulti_mecha}.



Let us begin by the proof the first part of the theorem:

\begin{lemma}
  A mechanical tree is strongly balanced. 
\end{lemma}

\begin{proof}
  Let $n,k\in\mathbb{N}$. For all node $v$, $h_v(n,k)$ is the
  number of $1$ in the factor of size $n$ and width $k$ rooted in $v$. We
  want to prove that for all pairs of nodes  $v$ and $v'$:
  $|h_v(n,k)-h_{v'}(n,k)|\leq 1$.  
  
  We assume that the nodes $v$ and $v'$ are inferior of phase $\phi$ and
  $\phi'$ (the proof with superior phases  is similar).
  \begin{equation*}
    h_v(n,k)-h_{v'}(n,k) = 
    \lfloor \frac{d^{n+k}-1}{d-1}\alpha+\phi\rfloor -
    \lfloor \frac{d^{k}-1}{d-1}\alpha+\phi\rfloor -
    \lfloor \frac{d^{n+k}-1}{d-1}\alpha+\phi'\rfloor +
    \lfloor \frac{d^{k}-1}{d-1}\alpha+\phi'\rfloor.
  \end{equation*}
  
  Using the well-known inequality  $ x-x'-1 < \lfloor x \rfloor - \lfloor x'
  \rfloor < x-x'+1$, one can show that
  \begin{equation*}
    -2<h_v(n,k)-h_{v'}(n,k)<2.
  \end{equation*}
  As $h_v(n,k)$ and $h_{v'}(n,k)$ are integers, we have $-1\leq
  h_v(n,k)-h_{v'}(n,)k \leq 1$ which ends the proof of the lemma. 
\end{proof}

We will see in the next section \ref{sec:link_sturmian} that a tree is
rational if and only if its density can be written as $\frac{p}{\Sn{n,k}}$
($p,k,n\in\mathbb{N}$), therefore we will do the proof of theorem
\ref{th:equiv_mecha_balanc} distinguishing strongly balanced tree with
density of this form or not.

\begin{lemma}
  If $\mathcal{A}$ is a strongly balanced tree of density $\alpha$ which
  can not be written as $\frac{p}{\Sn{n,k}}$ ($p,k,n\in\mathbb{N}$) then
  $\mathcal{A}$ is mechanical. 
\end{lemma}

\begin{proof}
  let $\tau$ be a real number and $v$ a node. At least one of the two
  following properties is true:
  \begin{eqnarray}
    \forall n\geq 1: h_v(n) \leq \lfloor \Sn{n}\alpha+\tau
    \rfloor,\label{eq:tau_1}\\
    \forall n\geq 1: h_v(n) \geq \lfloor \Sn{n}\alpha+\tau
      \rfloor. \label{eq:tau_2}
    \end{eqnarray}
    To prove this, assume that it is not true. Then there exists $k,n$ such
    that $h_v(n) < \lfloor \Sn{n}\alpha+\tau\rfloor $ and
    $h_v(k) > \lfloor \Sn{k}\alpha+\tau\rfloor$. In that case
    the number of $1$ in the factor of size $n,n-k$ (or $k,k-n$ if $k>n$)
    is $h_v(n) - h_v(k) \leq \lfloor\Sn{n}\alpha+\phi\rfloor
    - \lfloor \Sn{k}\alpha+\phi\rfloor -2 <
    \frac{d^{n}-d^k}{d-1}\alpha-1$ which violates the formula
    \eqref{eq:strong_bal}.
    
    Let us define now the number $\phi$ as the minimum $\tau$ that satisfies
    \eqref{eq:tau_1}
    \[\phi = \inf_\tau\Big\{\mathrm{For~all~}n: h_v(n) \leq \lfloor
    \Sn{n}\alpha+\tau \rfloor \Big\}.\] 
    
    For all $\tau>\phi$, the equation \eqref{eq:tau_1} is true, while for
    all $\tau'<\phi$, the equation \eqref{eq:tau_2} is true.  This means
    that for all $\epsilon>0$ and all $n$:
    \begin{equation}
      \Sn{n}\alpha+\phi-\epsilon-1 \leq \lfloor \Sn{n}\alpha+\phi-\epsilon \rfloor 
      \leq h_v(n) \leq \lfloor \Sn{n}\alpha+\phi+\epsilon \rfloor \leq \Sn{n}\alpha+\phi+\epsilon.
    \end{equation}
    Taking the  limit when $\epsilon$ tends to $0$ shows that:
    \begin{equation}
      \Sn{n}\alpha+\phi -1 \leq h_v(n) \leq
      \Sn{n}\alpha+\phi.
      \label{eq:tau_epsilon_0}
    \end{equation}
    Therefore, unless $\Sn{n}\alpha+\phi\in\mathbb{N}$,
    $h_v(n)=\lfloor \Sn{n}\alpha+\phi \rfloor = \lceil
    \Sn{n}\alpha+\phi-1 \rceil$. 
    
    If there exists $n\in\mathbb{N}$ such that
    $\Sn{n}\alpha+\phi\in\mathbb{N}$, then there are no other
    $k\in\mathbb{N}$ ($k\neq n$) such that
    $\Sn{k}\alpha+\phi\in\mathbb{N}$ -- otherwise that would
    violate the condition $\alpha \notin
    \{\frac{p}{\Sn{n,k}},p,k,q\in\mathbb{N}\}$. Therefore, for this
    particular $n$, either $h_v(n)=\Sn{n}\alpha+\phi=
    \lfloor \Sn{n}\alpha+\phi\rfloor$ -- in that case the node is
    inferior of phase $\phi$ -- or $h_v(n)= \Sn{n}\alpha+\phi-1
    = \lceil \Sn{n}\alpha+\phi-1 \rceil$ -- in that case
    the node is superior of phase $1-\phi$.
  \end{proof}


\begin{lemma}\label{mech}
Let ${\cal A}$  be a tree such that there exist $n$ and $k$ such that
all factors of size $(n,k)$ have
the same number of nodes with label 1. Then the tree is mechanical.
\end{lemma}

\begin{proof}
Let us take $n$ and $k$ satisfying the property, such that $n+k$ is
minimal and let us call $p$ is the common number of ones in the
factors of size $(n,k)$.
  Obviously, the tree as a density $\alpha =\frac{p(d-1)}{d^{k}(d^{n}-1)}$.

Let $v$ be the root of the tree.
 The same proof as in the irrational case can be used to establish
 that there exists $\phi$ such that 
  \[\Sn{n}\alpha+\phi -1\leq h_v(n)\leq\Sn{n}\alpha+\phi, \]
  and that the root  is inferior of phase $\phi$
  if there is  no $j$ such that $h_v(j)=\frac{d^j-1}{d-1}\alpha+\phi-1$ and
  superior of phase $1-\phi$ if there is no $i$ such that
  $h_v(i)=\frac{d^i-1}{d-1}\alpha+\phi$. Therefore the tree is mechanical
  unless there exist $i$ and $j$ satisfying these equalities. Let us show
  that if there exist such $i$ and $j$, there is a contradiction.
  
  Let $i=\min_{i'}\{ h_v(i')=\frac{d^{i'}-1}{d-1}\alpha+\phi \}$ and
  $j=\min_{j'}\{h_v(j')=\frac{d^{j'}-1}{d-1}\alpha+\phi-1\}$. Either $i<j$
  or $i>j$, let us assume that $j<i$, the other case is similar. The number
  of ones in the factor of size $i-j$ and width $j$ is
  $p'=\frac{d^i-d^j}{d-1}\alpha+1$. In that case we have $i\geq k+n$,
  otherwise that would violate the minimal property of $n+k$. If $j-i>n$ the factor of 
  size $i-j$ and width $j$ is composed of a factor of size $i-n$ and width
  $j$ and $d^{i-n-k}$ factors of size $n$ and width $k$ -- that have
  exactly $p$ nodes one as assumed in the previous paragraph -- and then
  the number of $1$ in this subtree is:   
  \[ h_v(i)-h_v(j) - d^{i-n-k}p+\phi+1 = \alpha\frac{d^{i-n}-d^{j}}{d-1}+\phi+1,\] 
  which violates the minimality of $i$.
  
  Then if all factors of size $(k,n)$ have exactly $p$ nodes 1, the tree is
  mechanical.
\end{proof}

\begin{lemma}\label{finite}
  If ${\cal A} $ is a strongly balanced tree with a density $\alpha =
  \frac{p}{\Sn{n,k}}$ then it has at most $n$ factors of size $n,k$
  with $p+1$ ones.
\end{lemma}

\begin{proof}
    Using Eq.  \eqref{eq:strong_bal}, each
  factor of size $(n,k)$ has $p-1$, $p$ or $p+1$ nodes labeled by $1$. As
  the tree is strongly balanced, either there is no factor with $p-1$
  ones or no factor with $p+1$ ones. Let us assume that there is no
  factor with $p-1$ ones (the
  other case is similar). We claim that there are at most $n$ factors of
  size $(n,k)$ with $p+1$ nodes labeled by $1$. 

  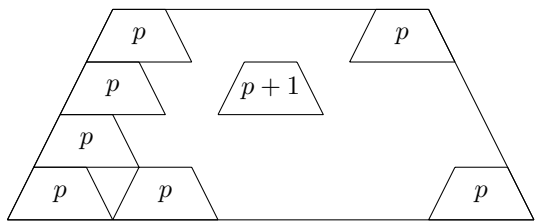
\begin{figure}[ht]
    \centering
   \begin{tikzpicture}[scale=0.7]
  \foreach \x/\y in {0/0,.5/1,1/2,1.5/3, 8/0,6.5/3, 2/0}
  \draw (\x-1,\y-0.5) -- (\x+1,\y-0.5)
  (\x+1,\y-0.5) -- (\x+0.5,\y+0.5)
  (\x+.5,\y+0.5) -- (\x-.5,\y+0.5)
  (\x-.5,\y+0.5) -- (\x-1,\y-0.5)
  node at (\x,\y) {$p$}; 

  \foreach \x/\y in {4/2}
  \draw (\x-1,\y-0.5) -- (\x+1,\y-0.5)
  (\x+1,\y-0.5) -- (\x+0.5,\y+0.5)
  (\x+.5,\y+0.5) -- (\x-.5,\y+0.5)
  (\x-.5,\y+0.5) -- (\x-1,\y-0.5)
  node at (\x,\y) {$p+1$}; 
  
  \draw (-1,-.5) -- (9,-.5)
  (9,-.5) -- (7,3.5)
  (7,3.5) -- (1,3.5)
  (1,3.5) -- (-1,-.5);
\end{tikzpicture}
    \caption{A block  of size $(n' = \ell n ,k')$ is made of $j$ blocks of
      size $(n,k)$.}
    \label{fig:bloc_n_k}
  \end{figure}
  
  Indeed, let $f$ be a factor of size $(n',k')$ with $n'=\ell n,
  i\in\mathbb{N}, k'\geq k$. This tree is composed of $j$ blocks of size
  $(n,k)$ (where $j$ depends on $\ell$ and $k'$, see Figure
  \ref{fig:bloc_n_k}) and using Eq. \eqref{eq:strong_bal} again, the number of nodes with label $1$ is either $jp-1$,
  $jp$ or $jp+1$. Therefore at most one of the $(n,k)$  blocks
  has $p+1$ nodes labeled by $1$. 

Now, in the whole tree, if  there were more than $n+1$ blocks of
  size $(n,k)$ with $p+1$ ones, each of these blocks starting at line
  $l_1,\dots$ and $l_{n+1}$, there would be two blocks with
  $l_i = l_j \mathrm{~mod~} n$ and the block of size $l_j-l_i+n, l_i$ would
  have $jp+2$ ones, which is not possible. Therefore there are at most $n$
  blocks of size $n,k$ with $p+1$ nodes labelled by $1$ in the whole tree.
\end{proof}

An example of a rational tree strongly balanced but not mechanical is
presented in Figure \ref{fig:ulti_mecha}.

\begin{lemma}
  A  strongly balanced tree  with density $\alpha = \frac{p}{\Sn{n,k}}$
  is ultimately mechanical. Furthermore, if the tree is irreducible, it is mechanical.
  \label{th:ratio_mecha}
\end{lemma}

\begin{proof}
  
Using Lemma \ref{finite},
    there are at most $n$ factors of size $n$ and
    width $k$ with $p+1$ nodes 1, in the rest of the tree all factors
    of size $(n,k)$ have exactly $p$ ones. Then the tree is ultimately
    mechanical by Lemma \ref{mech}.

 If the tree is irreducible, a factor appears either $0$ or an
    infinite number of times. As there are at most $n$ factors of size $(k,n)$
    with $p+1$ nodes 1, there are no such factors and the tree is
    mechanical by Lemma \ref{mech}.

Note that this lemma concludes the proof of Theorem \ref{th:equiv_mecha_balanc}.
 \end{proof}

\subsection{Link with Sturmian trees \label{sec:link_sturmian}}


In the case of words,  Sturmian word are exactly the
balanced (or mechanical) aperiodic words.  The case of trees does not work
as well since the Dyck Tree (Figure \ref{fig:dyck_tree}) and more generally
all examples of Sturmian trees given  in \cite{berstel2007fis} are not
balanced. However, the other  implication  holds as seen in  the
following theorem:

\begin{theorem}\label{th:mecha_is_ratio_sturmian}
The following propositions are true.
  \begin{itemize}
  \item A strongly balanced tree of density different from
    $\frac{p}{\Sn{n,k}}$ for any $p,n,k\in\mathbb{N}$ is Sturmian.
  \item A strongly balanced tree of density 
    $\frac{p}{\Sn{n,k}}$ for any $p,n,k\in\mathbb{N}$ is rational.
  \end{itemize}
\end{theorem}

This result has a simple implication: a strongly balanced tree is rational
if and only if there exist $p,n,k\in\N$ such that its density is
$\frac{p}{\Sn{n,k}}$. 

\begin{proof}
  Let us consider the case of inferior mechanical trees (the superior
  case being similar).
  
  Let $\mathcal{A}$ be a mechanical tree of density $\alpha$, let $v$ be a
  node and let $n\geq 0$.  According to Proposition \ref{prop:uniq},
  the factor of size $n$ only depends on the phase $\phi_v$ of its
  root. In fact, one can show that this factor only depends on the values
  $\lfloor\frac{d^i-1}{d-1}\alpha+\phi_v\rfloor$. For all $i\geq 0$ and
  $\phi\in[0:1]$, we define the quantities  $f_i(\phi) \bydef  \lfloor \frac{d^i-1}{d-1}\alpha+\phi \rfloor$.
  The number of factors of size $n$ only depends on the values
  $f_1(\phi),\ldots,f_{n}(\phi)$. 
  
  As seen in  \eqref{eq:all_phases}, the set of phases is dense in
  $[0;1]$, therefore they are exactly as many trees as tuples
  $f_1(\phi),\dots,f_{n}(\phi)$ when $\phi\in[0;1)$ by right-continuity of $f_i$.
  
  Each $f_i$ is an increasing functions taking integer values and
  $h_i(1)-h_i(0) = 1$. Thus there  are at most $n+1$ different tuples
  and  then at most $n+1$ factors of size $n$ and a mechanical tree is
  either rational or Sturmian. 
  
  Moreover if
  $\alpha\not\in\Big\{\frac{p}{d^k\Sn{n}}/p,n,k\in\mathbb{N}\Big\}$,
  we neither have $i\neq j$ and
  $\frac{d^i-1}{d-1}\alpha+\phi,\frac{d^j-1}{d-1}\alpha+\phi\in\mathbb{N}$
  and then there are  exactly $n+1$ factors of size $n$.

If $\alpha = \frac{p}{\Sn{n,k}}$, then the number of factors of size $n$ is
at most $n$ (see Section \ref{sec:all_phases}). Therefore the tree is rational using
Theorem \ref{th:rational}.
  
If the the tree is not mechanical, then Theorem
\ref{th:equiv_mecha_balanc}  says that the tree has density $\alpha =
\frac{p}{\Sn{n,k}}$ and  is ultimately
mechanical: There exists a depth $D\geq 1$ after which the tree is mechanical. 
Therefore, there are at most $\Sn{D}+n$ factors of any size
  ($n$ in the mechanical children because of the value of $\alpha$ plus
  $\Sn{D}$ in the prefix sub-tree). In that case the tree is rational
  by Theorem \ref{th:rational}.
\end{proof}

\section{Algorithmic issues \label{sec:algo}}

\subsection{Testing if a rational tree is strongly balanced}

Given a finite description of a rational tree, let us consider the problem of checking
whether this tree is balanced. An algorithm that  works in time
$0(n^3)$ where $n$ is the number of vertices of the minimal graph of the
tree is presented.

The first focus is  on the description of  the special structure of the minimal graph of a
rational strongly balanced tree. Then an algorithm for irreducible
rational trees is described  as well as a sketch of the algorithm for the general case. 

\subsubsection{Graph of rational strongly balanced trees\label{sec:graph_bal_tree}}

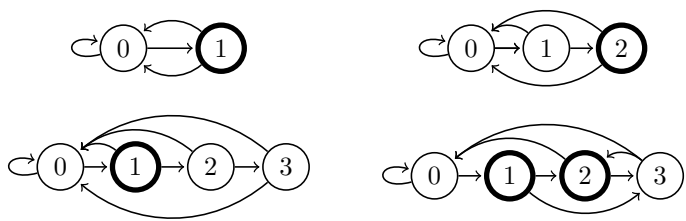
\begin{figure}[ht]
  \centering
  \begin{tabular}{ccc}
  \begin{tabular}{c}
    \begin{tikzpicture}[auto,node distance=1.3cm,semithick,shorten >=1pt,xscale=1,->]
      \tikzstyle{noeud}=[circle,draw];
      \node[noeud] (0) at (0,0) {0};
      \node[noeud,line width=2pt] (1) at (1.3,0) {1};
      
      \draw (0) edge (1)
       (0) edge[loop left] ()
      (1) edge[out=140,in=40] (0)
      (1) edge[out=-140,in=-40] (0);
    \end{tikzpicture}
  \end{tabular}&
  \begin{tabular}{c}
    \begin{tikzpicture}[auto,node distance=1.3cm,semithick,shorten >=1pt,xscale=1]
      
      \tikzstyle{noeud}=[circle,draw];
      \foreach \y in {0,1}
      \node[noeud] (\y) at (\y,0) {\y};
      \foreach \y in {2}
      \node[noeud,line width=2pt] (\y) at (\y,0) {\y};
      
      \foreach \x/\y in {0/1,1/2}
      \draw (\x) edge[->] (\y)
      (\y) edge[->,out=140,in=40] (0);
      
      \draw (0) edge[->] (1)
      (0) edge[loop left] ()
      (2) edge[->,out=-140,in=-40] (0);
    \end{tikzpicture}   
  \end{tabular}\\
  \begin{tabular}{c}
      \begin{tikzpicture}[auto,node distance=1.3cm,semithick,shorten >=1pt,xscale=1]
      
      \tikzstyle{noeud}=[circle,draw];
      \foreach \y in {0,2,3}
      \node[noeud] (\y) at (\y,0) {\y};
      \foreach \y in {1}
      \node[noeud,line width=2pt] (\y) at (\y,0) {\y};
      
      \foreach \x/\y in {1/2,2/3}
      \draw (\x) edge[->] (\y)
      (\x) edge[->,out=140,in=40] (0);
      
      \draw (0) edge[->] (1)
      (0) edge[loop left] ()
      (3) edge[->,out=140,in=40] (0)
      (3) edge[->,out=-140,in=-40] (0);
    \end{tikzpicture}  
  \end{tabular}
  &
  \begin{tabular}{c}
    \begin{tikzpicture}[auto,node distance=1.3cm,semithick,shorten >=1pt,xscale=1,->]
      
      \tikzstyle{noeud}=[circle,draw];
      \foreach \y in {0,3}
      \node[noeud] (\y) at (\y,0) {\y};
      \foreach \y in {1,2}
      \node[noeud,line width=2pt] (\y) at (\y,0) {\y};
      
      \foreach \x/\y in {0/1,1/2,2/3}
      \draw (\x) edge[->] (\y);
      
      \draw (0) edge[loop left] ()
      (1) edge[out=-40,in=-140] (3)
      (3) edge[out=140,in=40] (0)
      (3) edge[out=140,in=40] (2)
      (2) edge[out=140,in=40] (0);
    \end{tikzpicture}    
  \end{tabular}
\end{tabular}
  \caption{These graphs represent all mechanical trees of density $1/3$,
    $1/7$, $4/15$ and $6/15=2/5$. For all graphs with $n$ nodes, there are
    exactly $n$ different mechanical trees of this particular density,
    depending on which node is associated to the root. Note that the
    first three graphs have a very
    similar structure (Figure \ref{fig:irrat_mecha} displays more mechanical
    trees with this structure).} 
  \label{fig:mecha_ratio}
\end{figure}

Let us first consider  a rational mechanical tree of density $\alpha$. We
know that there exist $p,k,n\geq 0$ such that
$\alpha=\frac{p(d-1)}{d^{k}(d^n-1)}$. Using section \ref{sec:all_phases},
the  minimal graph has exactly $n+k$
nodes, and  for any node, the set of all possible phases of all its
descendants  is $[0;1)$. Therefore, the graph is strongly connected
and unique. The only difference between two rational mechanical trees of the same
density is to which node the root of the tree is associated.  Figure
\ref{fig:mecha_ratio} displays several examples. The (unique)  minimal
graph of the mechanical trees of density $1/3,1/7,4/15 $  and $2/15$ are
displayed.

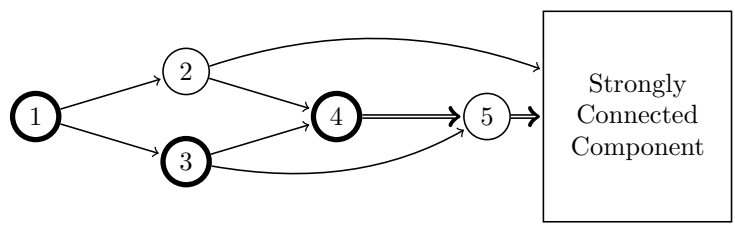
\begin{figure}[ht]
  \centering
  \begin{tikzpicture}[auto,node distance=1.3cm,semithick,->,shorten >=1pt, yscale=0.6]
  \tikzstyle{every node}=[draw];
  \node[unGraph,minimum size=3mm] at (0,0) (1) {1};
  \node[zeroGraph,minimum size=3mm] at (2,1) (2) {2};
  \node[unGraph,minimum size=3mm] at (2,-1) (3) {3};
  \node[unGraph,minimum size=3mm] at (4,0) (4) {4};
  \node[zeroGraph,minimum size=3mm] at (6,0) (5) {5};
  \node at (8,0) [minimum height=2.8cm] (F) {
    \begin{minipage}{.15\linewidth}
      \begin{center}
        Strongly\\Connected\\Component
      \end{center}
    \end{minipage}
  };

  \path (1) edge (2)
  (1) edge (3)
  (2) edge (4)
  (2) edge[bend left] (F)
  (3) edge (4)
  (3) edge[bend right] (5)
  (4) edge[double] (5)
  (5) edge[double] (F);
\end{tikzpicture}

  \caption{General form of the graph of a reducible strongly balanced tree:
    an acyclic graph ending in a unique strongly connected component.} 
  \label{fig:ulti_mecha_general_form}
\end{figure}

If the tree is strongly balanced but not mechanical, it is ultimately
mechanical (see proposition \ref{th:ratio_mecha}) which means that after a
finite depth $k$, all suffixes are mechanical trees with the same
density. All of these tree have the same graph,
therefore the minimal graph has a unique final strongly connected component which
is reached in at most $k$ steps. Therefore, the minimal graph of a strongly
balanced tree can be decomposed into  a finite acyclic graph and one
final strongly connected component, like in Figure
\ref{fig:ulti_mecha_general_form}. 

\subsubsection{Irreducible trees}

Testing if two graphs with  a given fixed out-degree are isomorphic  can be done in
polynomial time \cite{Luks}. Therefore using the result shown in the previous section
\ref{sec:graph_bal_tree}, an algorithm to test if a graph represents a
mechanical tree can be obtained  by computing the density $\alpha$ of the graph
and testing if the graph is isomorphic to the graph of all mechanical
trees with  density $\alpha$.  However this is not very efficient and here we propose
an algorithm  that  tests the balance property directly.
\medskip

Consider an irreducible rational tree $\mathcal{A}$ and
let  $n_0$ be  the number of vertices of its minimal graph. Theorem
\ref{th:mecha_is_ratio_sturmian} says that it is strongly balanced if and
only if it is mechanical. In that case its density is
$\frac{p}{\Sn{n_0,k_0}}$ for some $p,k_0\in\mathbb{N}$ and all sub-trees of
size $k_0,n_0$ have exactly $p$ nodes with label $1$.
Such factors will be called {\it basic blocks} in the following.

Recall that the tree is strongly balanced if all factors of size $(n,k)$ have
$\lfloor \alpha\Sn{n,k}\rfloor $ or $\lfloor
\alpha\Sn{n,k}+1\rfloor $ 
nodes of label one. We want to show that testing it for all $n,k <n_0+k_0$ is 
sufficient.

Let $v$ be a node and $n,k\geq 0$ and let $h_v(F)$ be the number of labels
$1$ in the factor $F$ of size $(n,k)$ with root $v$.

\begin{figure}[ht]
  \centering
  \begin{tikzpicture}[scale=.8]
  \newcommand\x{0}
  \draw (\x+0,2) -- (\x+-3,-4) -- (\x+3,-4) -- (\x+0,2);
  \draw (\x+-4,-2) -- (\x+3,-2);
  \draw[fill=black!10] (-3,-4) -- (3,-4) --  (2,-2) --
  (-2,-2) -- (-3,-4);
  
  \draw[dashed] (-1.25,-.5) -- (-1,-1);
  \draw (-1.5,-1) -- (-1.,-1) -- (-.5,-2);
  \node at (0,-1.5) {$\dots$};
  \draw[dashed] (1.25,-.5) -- (1,-1);
  \draw (1.5,-1) -- (1.,-1) -- (.5,-2);

  \draw (-3.5,2) edge[<->] node[fill=white] {$k$} (-3.5,-2)
  (-3.5,-2) edge[<->] node[fill=white] {$n$} (-3.5,-4);

  \draw (-2.3,-.5) edge[<->] node[fill=white,inner sep=2pt] {$k_0$} (-2.3,-1)
  (-2.3,-1) edge[<->] node[fill=white] {$n_0$} (-2.3,-2);
  \draw[dotted] (-2.5,-1) -- (5.5,-1);
  \draw[dotted] (-2.5,-.5) -- (2,-.5);
  \node at (-1.25,-1.5){p};
  \node at (1.25,-1.5){p};

  \node at (4,-1.5) {\Large $\mapsto$};

  \renewcommand\x{8}
  \draw (\x+0,2) -- (\x+-3,-4) -- (\x+3,-4) -- (\x+0,2);
  \draw (\x+-2.5,-1) -- (\x+4,-1);
  \draw[fill=black!10] (\x+-3,-4) -- (\x+3,-4) --  (\x+1.5,-1) --
  (\x+-1.5,-1) -- (\x+-3,-4);
  
  \draw (\x+3.5,2) edge[<->] node[fill=white] {$k-n_0$} (\x+3.5,-1)
  (\x+3.5,-1) edge[<->] node[fill=white] {$n+n_0$} (\x+3.5,-4);
  

  \node[fill=white, rounded corners] at (0,-3) {$h(n,k)$};
  \node[fill=white, rounded corners] at (\x+0,-2.5) {$h(n,k)+T_{k}$};
\end{tikzpicture}
  \caption{The first transformation: if $k>n_0+k_0$, we add a level of
    factors of size $n_0,k_0$ that all contain exactly $p$ ones. The size
    of the factor becomes $(n+n_0,k-n_0)$. We repeat
    the transformation until the size is $(n',k')$ with $k'<n_0+k_0$.
In the figure, $T_{k}$ stands for $pd^{k-n_0-k_0} p$.} 
  \label{fig:adding_subtrees}
\end{figure}
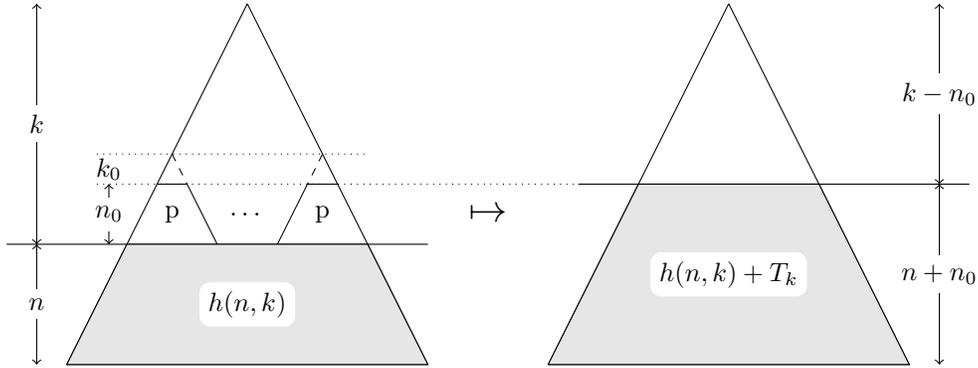

Starting from $F$, we construct a new factor $F'$ by adding a new  factor on top of $F$ of size  $n_0,k-n_0$. 
This new factor can be  partitioned into  $d^{k-n_0-k_0} $ basic
blocks. 
The total factor $F'$ is of size $(n+n_0, k-n_0)$ and its  number of
ones is $h_v(F') = h_v(F) + d^{k-n_0-k_0} p$ (see Figure
\ref{fig:adding_subtrees}).


The augmentation of the factor can be repeated until its size $n',k'$ is such that
$k' \leq k_0+n_0$. Its  number of ones is $h_v(F') = h_v(F) + H$ where $H$ does not depend
on $v$.

\begin{figure}[ht]
  \centering
  \begin{tikzpicture}[scale=.8]
  \newcommand\x{0}
  \draw[fill=black!10] (\x+-3,-4) -- (\x+3,-4) --  (\x+1,0) --
  (\x+-1,0) -- (\x+-3,-4);
  \draw (\x+0,2) -- (\x+-3,-4) -- (\x+3,-4) -- (\x+0,2);
  \draw (\x+-4,0) -- (\x+2,0);
  
  \node[fill=white, rounded corners] at (\x+0,-1.75) {$h(n',k')$};
  
  \draw (-3,-2.5) edge[<->] node[fill=white,inner sep=2pt] {$k_0$} (-3,-3)
  (-3,-3) edge[<->] node[fill=white,inner sep=2pt] {$n_0$} (-3,-4);

  \draw (-3.5,2) edge[<->] node[fill=white] {$k'$} (-3.5,0)
  (-3.5,0) edge[<->] node[fill=white] {$n'$} (-3.5,-4);

  \draw[dashed] (-2.25,-2.5) -- (-2,-3);
  \draw (-2.5,-3) -- (-2,-3) -- (-1.5,-4);
  \node at (0,-3.5) {$\dots$};
  \draw[dashed] (2.25,-2.5) -- (2,-3);
  \draw (2.5,-3) -- (2,-3) -- (1.5,-4);

  \draw[dotted] (-3,-3) -- (5,-3);
  \draw[dotted] (-3,-2.5) -- (3,-2.5);
  \node at (-2.25,-3.5){p};
  \node at (2.25,-3.5){p};
    
  \node at (4,-1.5) {\Large $\mapsto$};

  \renewcommand\x{8}
  \draw (\x+0,2) -- (\x+-3,-4) -- (\x+3,-4) -- (\x+0,2);
  \draw (\x+-2,0) -- (\x+4,0);
  \draw[fill=black!10] (\x+-2.5,-3) -- (\x+2.5,-3) --  (\x+1,0) --
  (\x+-1,0) -- (\x+-2.5,-3);
  
  \draw (\x+3.5,2) edge[<->] node[fill=white] {$k'$} (\x+3.5,0)
  (\x+3.5,0) edge[<->] node[fill=white] {$n'-n_0$} (\x+3.5,-3);
  
  \draw (5,-3) -- (\x+4,-3);

  \node[fill=white, rounded corners] at (\x+0,-1.5)
  {$h(n',k'){-}T_{n'}$}; 
\end{tikzpicture}
   
  \caption{The second transformation: if $n'>n_0+k_0$, we can remove a level
    of factors of size $(n_0,k_0)$. The size of the factor becomes
    $(n'-n_0,k')$. We repeat the transformation until the size is $(n',k')$
    with $n'<n_0+k_0$ (here, $T_{n'}  =  p d^{n'-n_0-k_0}$).}
  \label{fig:remove_subtrees}
\end{figure}
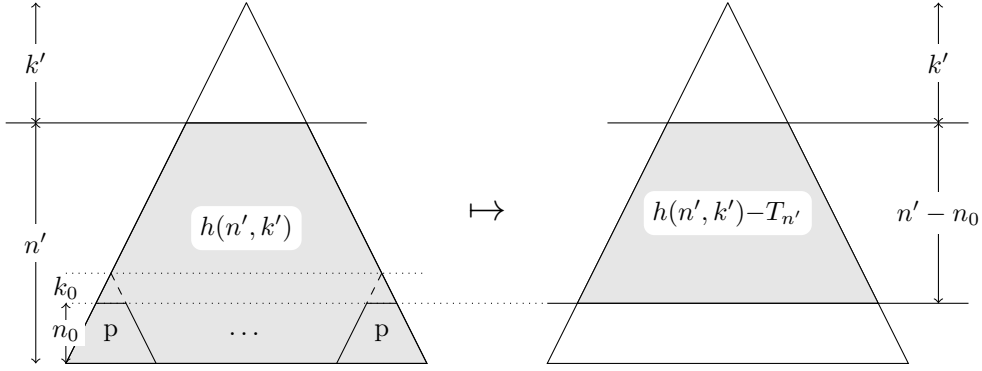

The second phase consists in building a new factor $F''$ by removing a factor from $F'$ of size $n_0,
k'+n'-n_0$. The removed part can be partitioned  into $ d^{n'-n_0-k_0}$
basic blocks.
Therefore  the number of ones in  $F''$ is $h_v(F'') = h_v(F') -
d^{n'-n_0-k_0}p$. This transformation is illustrated in Figure \ref{fig:remove_subtrees}.

By repeating this transformation as long as $n'' > n_0+k_0$,
we get a final factor $F''$ whose size is $(n'',k'') $ with  $n''
<n_0+ k_0$, $k'' < n_0+ k_0$  
and whose number of ones is $h_v(F'') = h_v(F) + H -K$, where $H$ and
$K$ do not depend on $v$ but only on $n$ and $k$.

Since $h_v(F) = h_v(F'') - H + K$, it is enough to compute the number
of ones in all factors with size $(n'',k'') $ where  $n''
<n_0+ k_0$, $k'' < n_0+ k_0$,  to be able to obtain the number of ones
in all factors on any size.

Also, it is enough to test if all factors  with size $(n'',k'') $ where  $n''
<n_0+ k_0$, $k'' < n_0+ k_0$  satisfy the strong balance property for
all factors on any size to have the same property.

 There are at most $n$ sub-trees of a given height and width.  For
 $\ell <m$,
 let us call $h_{i,\ell,m}$ the number of $1$ in the $i^{\mathrm{th}}$
 sub-tree of
 height $\ell$ and width $\ell+m$. Let us call $v(i)=(v_1(i),\dots,v_d(i))$ the set
 of the $d$ children of the tree $i$. $h_{i,\ell,m}$ can be computed using the
 formula: 

 \begin{equation}
   h_{i,\ell,m} = \left\{
     \begin{array}{lll}
       h_{i,1,0} &=& \gamma_i\\
       h_{i,\ell,0} &=& \gamma_i + \sum_{j\in v(i)} h_{j,\ell-1,0}\\
       h_{i,\ell,m} &=& \sum_{j\in v(i)} h_{j,\ell-1,m-1} 
     \end{array}
   \right.\label{eq:h_i_l_m}
 \end{equation}

These considerations yield the following
algorithm \ref{algo:test_strbal}.

\begin{algorithm}
  \caption{Testing if a rational tree is strongly balanced}
  \label{algo:test_strbal}
  \begin{algorithmic}
    \REQUIRE Minimal graph $G$ of a irreducible rational tree
    \ENSURE The tree corresponding to $G$ is strongly balanced
    \STATE N:= number of vertices of $G$\;
    \STATE Compute the density $\alpha$ of the Markov Chain\;
    \IF{for all $k$:$\frac{d^N-d^k}{d-1}\alpha\not\in\mathbb{N}$}
    \RETURN ``not strongly balanced''\;
    \ENDIF

    \FOR{$1\leq i,n,k\leq N$ } 
    \STATE Compute $h_{i,n,k}$ according to \eqref{eq:h_i_l_m}\;
    \IF {$h_{i,n,k} \not=  \lfloor \frac{d^n-d^k}{d-1}\alpha \rfloor$ and 
      $h_{i,n,k} \not=  \lfloor \frac{d^n-d^k}{d-1}\alpha \rfloor+1$}
    \RETURN ``not strongly balanced''\;
    \ENDIF
    \ENDFOR
    
    \RETURN ``strongly balanced''\;
  \end{algorithmic}
\end{algorithm}

Solving the Markov chain to get $\alpha$ takes at most $O(N^3)$
operations. Writing  the density under the form $\frac{p}{d^N-d^k}$ is linear in $N$ and
computing all $h_{i,\ell,m}$ takes $0(N^3)$ operations using the formula
\eqref{eq:h_i_l_m}. Therefore the algorithm runs in time $O(N^3)$. 

\subsubsection{General case}

The general case is more complicated since there can be some factors of
size $(n_0,k_0)$ with $p+1$ (or $p-1$) nodes labeled by $1$. However
the structure of  the minimal graph
of strongly balanced trees made in Section \ref{sec:graph_bal_tree}
can be useful.
\begin{itemize}
\item Indeed, the minimal graph must have only one strongly connected component and
  it must corresponds to a strongly balanced tree.
\item If the density of the strongly connected component is $\frac{p}{2^{n_0}C{k_0}}$, all
  factors of size $n_0,k_0$ in the strongly component have exactly $p$ nodes
  labeled by $1$. 
\end{itemize}

Therefore, using the same techniques of reduction of the size as in Figure
\ref{fig:adding_subtrees}, one can show that we just have to test the
balanced property for factors of size at most $(n,n)$ where $n$ is the number
of vertices in the graph.

\subsection{Counting}

In this part, we address the problem of counting all possible factors of a
mechanical tree. We will focus on trees of degree $2$ and will compare this
to the total number of possible factors of  binary trees.

There are $2^n$ finite words of length $n$. Not all these words can be
factors of a Sturmian words -- for example $0011$ can not be since it is not
balanced. In fact, the number of factors of length $n$ of Sturmian words is 
\begin{equation}
  1+\sum_{i=1}^m(m-i+1)\phi(i)\label{eq:number_words}
\end{equation}
where $\phi$ is the Euler function -- $\phi(i)$ is the number of integers
less than $i$ and coprime with $i$. Asymptotically, this number is
equal to $m^3/\pi^2$. 

The number $a_n$ of unordered complete binary trees of height $n$ satisfies the equation: 
\begin{equation}
  \label{eq:recurence_unordered}
  a_{n+1} = a_n(a_n+1)
\end{equation}

According to \cite{sloane2003lei}, there is no simple solution of this
equation but using the method described in \cite{aho1973sde}, one can show that $a_n$ is
the nearest integer close to $\theta^{2^n}-1/2$, where $\theta \approx
1.597910218$ is the exponential of the rapidly convergent series
$\ln(3/2)+\sum_{n \geq 0} \ln(1+(2a_n+1)^{-2})$.

In section \ref{sec:link_sturmian}, we have seen that the number of factors
of size $n$ of a Sturmian tree is the number of tuple
$(f_1(\phi,\alpha),\dots,f_n(\phi,\alpha))$ where
$f_i(\phi,\alpha)=\lfloor(2^n-1)\alpha+\phi\rfloor$.  Let us call $u_n$
this number. 
\begin{figure}[ht]
  \centering
  \caption{Lines $\phi=(2^n-1)\alpha - i$ for $n\geq 0$ and $0\leq i<2^n-1$}
  \label{fig:counting}
\end{figure}
To count the number of these tuples, we draw the lines for which
$(2^n-1)\alpha-\phi\in\mathbb{N}$, ( see Figure \ref{fig:counting}). The
number of tuples is the number of different zones on this figure.

An exact count $u_n$ is cumbersome to obtain but good bounds can be computed easily.
$u_{n+1}-u_n$ corresponds to the number of zones
added by the adding the  lines $\alpha\mapsto(2^{n+1}-1)\alpha-i$. Each of
these $2^{n+1}-1$ lines:
\begin{itemize}
\item at least add a new zone if it only crosses other lines at points $\phi=0$
  or $\phi=1$. This is a very low estimate since it is only true for $i=0$
  or $i=2^n-2$, in the other cases it  crosses at least the line
  $\alpha\mapsto\phi$.
\item at most add $1+n$ zones if it crosses the $n$ lines corresponding to
  $\alpha\mapsto (2^j-1)\alpha-i_j$, $1\leq j\leq n$ and if all these points are
  pairwise distinct. 
\end{itemize}

Therefore we have an estimation for all $n\geq 2$:
\begin{equation}
  \label{eq:counting_estimate_1}
  2+ 2(2^{n+1}-3) \leq u_{n+1}-u_n \leq (n+1)(2^{n+1}-1)
\end{equation}
This leads to the bounds for $n\geq 3$:
\begin{equation}
  \label{eq:counting_estimate}
  2^n\leq u_n \leq (n-1)u_n.
\end{equation}

Improving these bounds seems difficult. To do so, one would have to count
whether a ``new'' intersection has already been counted or if it is on the
boundary $\phi=0$. By simulation, it seams that the
number of trees is closer to $n2^{n}$ than to $2^n$. 

\section{Extremal properties}

In this section, we show that strongly balanced trees are extremal for
certain convex cost functions that can be used in scheduling problems.

  Let $g:\mathbb{R}^+ \rightarrow \mathbb{R}^+$ be a convex function. Let us
  assume that $g$ has a minimum in $0 < \alpha < 1$.

 For each node
  $v$ and each factor of size $n$ rooted in $v$, , we define a cost of
  the factor rooted in $v$ as a convex function of the  density of
  ones: $d_v(n) = h_v(n) / S(n)$:

  $$ C_v(n) \bydef  g(d_v(n)).$$

  We can define a cost $C^k$ of order $k$ of the total tree by
  considering all nodes in the sub tree of height $\ell$ rooted in
  $r$, ${\cal A}_\ell$ as the
  Cezaro limit:

  $$ C^k\bydef  \limsup_{\ell \to \infty} \frac{\sum_{v\in{\cal A}_\ell}
    C_v(k)}{S(\ell)}.$$ 

  For each $k$, this cost is minimized when the number of $1$ in a tree of
  height $k$ is between $~~ S(k)~$ and
  $~~ S(k)~$. This  means that a strongly balanced tree will minimize
  any increasing function of all $C_k$ (for example the average value
  over all $k$).

This has potential applications
  in optimization problem in distributed systems with a binary causal
  structure and would generalize some  results in \cite{AGHBook}.

Consider for example a scheduling problem with  two processors (with
related speeds $u_0$ and $u_1$) and an infinite  set of tasks with a
dependency pattern for forms a  tree with degree $d$.
Tasks at level $k$ in the tree have size $1/d^k$ and are released 1
unit of time after their father.

\begin{theorem}
Under the ongoing assumptions,
there  exists an optimal density $\alpha \in (0,1)$ such that a strongly
balanced tree with density $\alpha$ is the optimal allocation of tasks
to processors $0$ and $1$, in terms of average flowtime.
\end{theorem}

\begin{proof}(sketch)
First, one should notice that at each second a load of one unit of
work arrives in the system.
Also note that every second, the allocation pattern forms a balanced
sequence prefix with density $\alpha$. 
Finally, up to level $k$, all tasks can be seen as clusters of tasks of size
$1/d^k$.

In \cite{AGHBook}, it is shown that the optimal allocation (for the
average flow time) when tasks come in 
clusters of size $m$ (for any $m$) is a balanced sequence over the clusters.

Using a diagonal process over all sizes of clusters  that up to
level $k$ show that strongly balanced trees are optimal up to level
$k$.
This is trus for any $k$.

The end of the proof comes from taking the limit when $k$ goes to infinity.
\end{proof}

Note that an arrival pattern that forms a tree of degree $d$ may arise when tasks 
are generated by a recursive program.
Actually, this result can be generalized to more general arrival patterns.
If tasks at level $k$ in the tree are released  after iid stochastic
times (with an arbitrary distribution but expectation equal to one),
yet again an allocation of the tasks to processors 0 and 1 according to a  strongly balanced
tree is optimal, however the proof of this result is beyond the scope
of this paper.

\section{Glossary \label{sec:glossary}}

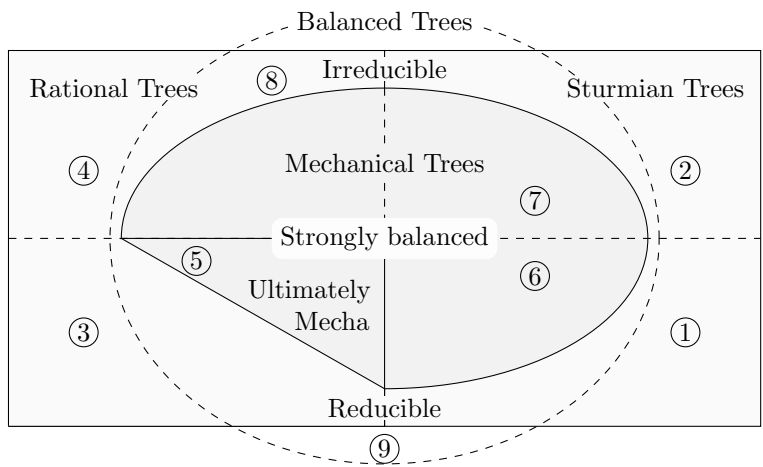
\begin{figure}[ht]
  \centering
  \begin{tikzpicture}
  \draw[fill=black!2
  ] (0,0) -- (5,0) -- (5,5) -- (0,5) -- (0,0);
  \draw[fill=black!2] (5,0) -- (10,0) -- (10,5) -- (5,5) -- (5,0);
  \node at (5,2.5) [ellipse, fill=black!5, minimum width=7cm, minimum
  height=4cm,draw=black] {}; 
  \draw[fill=black!2] (1.5,2.5) -- (5,0.5) -- (5,0) -- (0,0);
  \draw (1.5,2.5) -- (5,2.5) -- (5,0.5);
  \draw[dashed] (5,2.5) -- (5,4.5) (0,2.5) -- (10,2.5);

  \node[ellipse,draw,minimum height=6cm, minimum width=7.3cm,dashed] at
  (5,2.5) {}
  node[fill=white, rounded corners] at (5,5.4) {Balanced Trees};
  
  \node at (5,3.5) [fill=black!5] {Mechanical Trees}
  node at (4,1.8) {Ultimately}
  node at (4.3,1.4) {Mecha}
  node[fill=black!1, rounded corners] at (5,2.5) {Strongly balanced};
  
  \node[fill=black!2,inner sep=1pt] at (1.4,4.5) {Rational Trees}
  node[fill=black!2,inner sep=1pt] at (5,4.75) {Irreducible}
  node[fill=black!2,inner sep=1pt] at (5,0.25) {Reducible};
  \node[fill=black!2,inner sep=1pt] at (8.6, 4.5) {Sturmian Trees};
  
  \tikzstyle{ctr_example}=[circle,draw,inner sep=1pt]
  \node[ctr_example] at (2.5,2.2) {\ref{str_bal_not_mecha}};
  \node[ctr_example] at (5,-.3) {\ref{irratbal_not_str_bal}};
  \node[ctr_example] at (3.5,4.6) {\ref{ratbal_not_str_bal}};
  \node[ctr_example] at (9,1.25) {\ref{sturmian_not_bal}};
  \node[ctr_example] at (9,3.4) {\ref{irred_sturmian_not_bal}};
  \node[ctr_example] at (1,1.25) {\ref{irreducible_tree_not_bal}};
  \node[ctr_example] at (1,3.4) {\ref{reductible_tree_not_bal}};
  \node[ctr_example] at (7,3) {\ref{irred_mecha}};
  \node[ctr_example] at (7,2) {\ref{redu_mecha}};

\end{tikzpicture}
  \caption{Relations of inclusion linking the different definitions that we
    presented. Each number refers to an example detailed in section
    \ref{sec:glossary}. For example \ref{str_bal_not_mecha} is the set
    of  trees that are rational, reducible, ultimately mechanical, strongly
    balanced, balanced and neither mechanical nor Sturmian.} 
  \label{fig:diagramme}
\end{figure}

The aim of this part is to give the big picture and to provide several examples of trees that are
either balanced, strongly balanced, reducible, irreducible, rational or
Sturmian. In particular, we will give counter-examples that shows that the
inclusions between these notions are strict. The Figure
\ref{fig:diagramme} illustrates these results. 

\begin{enumerate}
\item \emph{Reducible Sturmian tree that is not balanced} \label{sturmian_not_bal} --
  contrarily to the case of words where Sturmian words are balanced, there
  exist Sturmian trees that are not balanced. The Dyck tree, see
  Figure \ref{fig:dyck_tree}, is one of them.

\item \label{irred_sturmian_not_bal} \emph{Irreducible Sturmian trees  that
    are not balanced} -- An example of a Sturmian tree that is
  irreducible (but not balanced) is the \emph{reflected random walk} tree
  represented in Figure \ref{fig:dyck2}. It is Sturmian since the equivalence
  classes of the relation $\equiv_n$ are
  $\{0\},\dots,\{n-1\},\{n,n+1,\dots\}$. 
  \begin{figure}[ht]
    \centering
    \begin{tabular}{cc}
      \begin{tabular}{c}
        \begin{tikzpicture}[yscale=0.6]
  \node[un]{}
  child {node[zero]{}
    child {node[zero]{}
      child {node[zero]{}
        child {node[zero]{} child[pointille] }
        child {node[zero]{} child[pointille] }}
      child {node[zero]{}
        child {node[zero]{} child[pointille] }
        child {node[un]{} child[pointille] }}}
    child {node[un]{}
      child {node[zero]{}
        child {node[zero]{} child[pointille] }
        child {node[un]{} child[pointille] }}
      child {node[un]{}
        child {node[zero]{} child[pointille] }
        child {node[un]{} child[pointille] }}}}
  child {node[un]{}
    child {node[zero]{}
      child {node[zero]{}
        child {node[zero]{} child[pointille] }
        child {node[zero]{} child[pointille] }}
      child {node[un]{}
        child {node[zero]{} child[pointille] }
        child {node[un]{} child[pointille] }}}
    child {node[un]{}
      child {node[zero]{}
        child {node[zero]{} child[pointille] }
        child {node[zero]{} child[pointille] }}
      child {node[un]{}
        child {node[zero]{} child[pointille] }
        child {node[un]{} child[pointille] }}}};
\end{tikzpicture}
  
      \end{tabular}
      &
      \begin{tabular}{c}
        \begin{tikzpicture}[auto,node distance=1.3cm,semithick,shorten >=1pt,yscale=0.6,xscale=0.7]

  \tikzstyle{noeud}=[circle,draw];
  \foreach \x/\y in {1,...,5}
  \node[noeud] (\y) at (1.3*\y,0) {\y};
  
  \node[noeud,line width=2pt] (0) at (0,0) {0};
  
  \node (ENTREE) at(0,1.5) {};
  \node[circle,pointille] (6) at (1.3*6,0) {...};
  
  \foreach \x/\y in {0/1,1/2,2/3,3/4,4/5,5/6}
  \draw (\y) edge[->,bend left] (\x)
  (\x) edge[->,bend left] (\y);
  
  \draw (0) edge[loop below,->]()
  (ENTREE) edge[->] (0);

\end{tikzpicture}

      \end{tabular}
    \end{tabular}
    \caption{The reflected random walk tree: each node of type $n$ is
      followed by one of type $n-1$ and one of type $n+1$ (except for $0$
      that is followed by $0$ and $1$). } 
    \label{fig:dyck2}
  \end{figure}
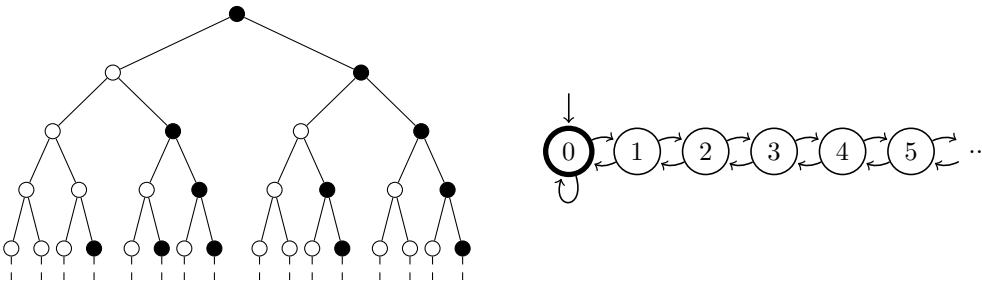

\item \label{irreducible_tree_not_bal} \emph{Irreducible rational trees} --
  see Figure \ref{fig:density_example}.

\item \label{reductible_tree_not_bal} \emph{Reducible rational trees} --
  see Figure \ref{fig:rat}.
  
\item \emph{Rational reducible strongly balanced tree that is not 
    mechanical} \label{str_bal_not_mecha} -- strongly balanced tree are not
  necessarily mechanical in the case of reducible rational trees but only
  ultimately mechanical, see Figure \ref{fig:ulti_mecha} for an example. 
  
\item\label{redu_mecha} \emph{Reducible mechanical trees} -- let
  $\alpha$ be a normal number and consider the mechanical tree of density
  $\alpha$ and phase $0$ at the root. As $\alpha$ is normal, there is a unique
  phase corresponding to each node of order $k$ which is the fractional
  part of: 
  \begin{equation}
    \label{ref:phases_phi_0}
    \alpha(\frac{1}{d^k}+\dots+\frac{1}{d})+\frac{i_k}{d^k}+\dots+\frac{i_1}{d} 
  \end{equation}
  for a unique sequence $i_1,\dots,i_k$. 
  One can show that if two sequences of $i_1,\dots,i_k$ are different, then
  these phases are different, which shows that the minimal graph of the
  tree is exactly the tree itself. 
  
\item \label{irred_mecha} \emph{Irreducible mechanical trees} --
  let $w$ be a mechanical word and consider a graph with vertices
  $\{0,1,\dots,\}$, where a node $i\geq 0$ has label one if and only if
  $w_i=1$. The node $i$ has two outgoing arcs: one ending in $i+1$, one
  ending in $0$. We call this graph a \emph{restart tree} since for a node
  $n$, we have the choice between restarting back in $0$ or continuing in
  $n+1$, an example is displayed in Figure \ref{fig:irrat_mecha}. 

  \begin{figure}[ht]
    \centering
    \begin{tikzpicture}[auto,node distance=1.3cm,semithick,shorten >=1pt,yscale=0.6,xscale=0.7]

  \tikzstyle{noeud}=[circle,draw];
  \foreach \x/\y in {0,1,3,4,5}
  \node[noeud] (\y) at (1.3*\y,0) {\y};
  \foreach \x/\y in {2,6}
  \node[noeud,line width=2pt] (\y) at (1.3*\y,0) {\y};
  
  \node[circle,pointille] (7) at (1.3*7,0) {...};
  
  \foreach \x/\y in {1/2,2/3,3/4,4/5,5/6,6/7}
  \draw (\x) edge[->] (\y)
  (\x) edge[->,out=140,in=40] (0);

  \draw (0) edge[->] (1)
  (0) edge[loop left] ();

\end{tikzpicture}

    \caption{Example of the restart  tree corresponding to the word
      $aabaaab\dots$}
    \label{fig:irrat_mecha}
  \end{figure}
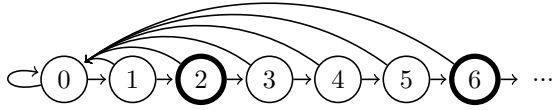

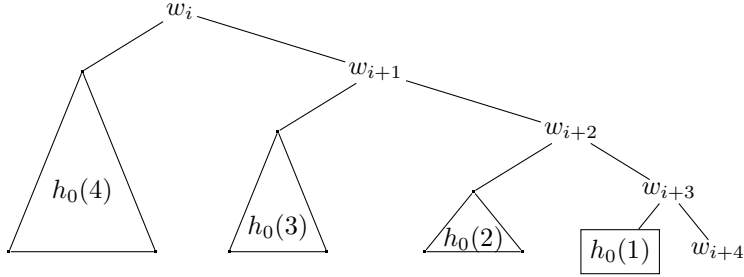
\begin{figure}[ht]
  \centering
  \begin{tikzpicture}[xscale=1.3, yscale=0.8]
  \tikzstyle{sommet}=[inner sep=1pt]
  \tikzstyle{arbre}=[inner sep=0pt,fill]
  
  \node[sommet] at (0,0) (V1) {$w_i$}
  node[sommet] at (2,-1) (V2) {$w_{i{+}1}$}
  node[sommet] at (4,-2) (V3) {$w_{i{+}2}$}
  node[sommet] at (5,-3) (V4) {$w_{i{+}3}$}
  node[sommet] at (5.5,-4) (V5) {$w_{i{+}4}$}
  node[arbre] at (-1,-1) (s11) {}
  node[arbre] at (-1.75,-4) (s12) {}
  node[arbre] at (-0.25,-4) (s13) {}
  node at (-1,-3) {$h_{0}(4)$}
  node[arbre] at (1,-2) (s21) {}
  node[arbre] at (.5,-4) (s22) {}
  node[arbre] at (1.5,-4) (s23) {}
  node at (1,-3.6) {$h_{0}(3)$}
  node[arbre] at (3,-3) (s31) {}
  node[arbre] at (2.5,-4) (s32) {}
  node[arbre] at (3.5,-4) (s33) {}
  node at (3,-3.8) {$h_{0}(2)$}
  node[rectangle,draw] at (4.5,-4) (s41) {$h_0(1)$};
  
  \path (V1) edge (V2)
  (V1) edge (s11)
  (V2) edge (V3)
  (V2) edge (s21)
  (V3) edge (V4)
  (V3) edge (s31)
  (V4) edge (V5)
  (V4) edge (s41);
    
  \path (s11) edge (s12) 
  (s12) edge (s13) 
  (s13) edge (s11)
  (s21) edge (s22) 
  (s22) edge (s23) 
  (s23) edge (s21)
  (s31) edge (s32) 
  (s32) edge (s33) 
  (s33) edge (s31)
  ;
\end{tikzpicture}
  \caption{Number of ones in a factor of the restart tree of size
    $5$} 
  \label{fig:number_irrat}
\end{figure}

As seen in Figure \ref{fig:number_irrat}, the number of
ones in a factor of size $n$ that corresponds to the node $i$ is  
\begin{equation}
  h_i(n) = w_i+\dots+w_{i+n-1} + h_0(n-1) + \dots + h_0(1),
\end{equation}
and the number of ones in a factor of size $n$ and width $k$ is 
\begin{equation}
  h_i(n,k) = h_i(n)-h_i(k) = w_k+\dots+w_{i+n-1} + h_0(n-1) + \dots +
  h_0(k). 
\end{equation}
Therefore the tree is strongly balanced if and only if the word $w$ is
balanced. Since the tree is irreducible, in that case the tree is also
mechanical. Moreover we can show that for any word $w$ the tree has a
density which is $\lim_{n\to\infty} \frac{h_0(n)}{2^n-1} = \frac{w_0}{2} +
\frac{w_1}{4}+\frac{w_2}{8}+\cdots$.

Thus for any aperiodic balanced word, this gives us an example of
irreducible irrational strongly balanced tree. 

\item \label{ratbal_not_str_bal} \emph{Rational balanced tree that is not strongly balanced} --
  An example of rational trees balanced but not strongly balanced is
  presented in Figure \ref{fig:ratio_bal_not_strbal}.  On can show that all
  of its factors of size $3$ have exactly $4$ nodes of label one. Using
  this fact, one can show that the number of ones in a factor of size
  $3n+i$ ($0\leq i\leq 3$) rooted in a node $j$ is: 
  
  \begin{center}
    \begin{tabular}{|l|cccc|}
      \hline
      Size & Node 1 & Node 2 & Node 3 & Node 4\\
      \hline
      $3n$&$4\frac{8^n-1}{7}$&$4\frac{8^n-1}{7}$&$4\frac{8^n-1}{7}$&$4\frac{8^n-1}{7}$\\
      $3n+1$&1+$2.4\frac{8^n-1}{7}$&$0+2.4\frac{8^n-1}{7}$&$0+2.4\frac{8^n-1}{7}$&1+$2.4\frac{8^n-1}{7}$\\
      $3n+2$&1+$4.4\frac{8^n-1}{7}$&$1+4.4\frac{8^n-1}{7}$&$2+4.4\frac{8^n-1}{7}$&2+$4.4\frac{8^n-1}{7}$\\
      \hline
    \end{tabular}
  \end{center}
  
  This shows that the tree is balanced. It is not strongly balanced since
  there are factors of size $(1,1)$ with $2$ nodes labeled by one
  and others with $0$ nodes labeled by one as seen in the bottom right part
  of figure \ref{fig:ratio_bal_not_strbal}. Also its minimal graph is not
  isomorphic to the unique minimal graph of a mechanical tree of density
  $4/7$ that has only $3$ nodes (see the discussion about graphs of
  strongly balanced tree section \ref{sec:graph_bal_tree}).
  
  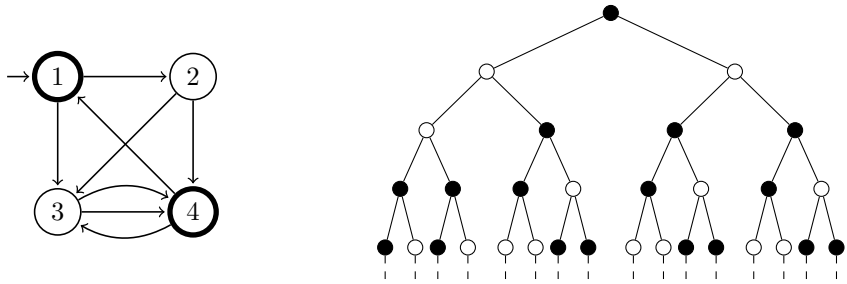
\begin{figure}[ht]
    \centering
    \begin{tabular}{ll}
      \begin{minipage}{0.3\linewidth}
        \begin{tikzpicture}[->,auto,node distance=1.8cm,shorten >=1pt,semithick]
  
  \tikzstyle{Anode} = [unGraph]
  \tikzstyle{Bnode} = [zeroGraph]
  \tikzstyle{Cnode} = [Bnode]
  \tikzstyle{Dnode} = [Anode]
  
  \node (FIRST) {};
  \node (A) [Anode,right of=FIRST, node distance=.8cm] {1};
  \node[Bnode] (B) [right of=A] {2};
  \node[Cnode] (C) [below of=A] {3};
  \node[Dnode] (D) [right of=C] {4};
  
  \path (FIRST) edge (A);
  \path (A) edge (B)
  (A) edge (C)
  (B) edge (D)
  (B) edge (C)
  (C) edge (D)
  (C) edge [bend left] (D)
  (D) edge[bend left] (C)
  (D) edge (A);
\end{tikzpicture}
      \end{minipage}
      &
      \begin{tabular}{c}
        \begin{tikzpicture}[yscale=0.6]
  \tikzstyle{Anode} = [un]
  \tikzstyle{Bnode} = [zero]
  \tikzstyle{Cnode} = [Bnode]
  \tikzstyle{Dnode} = [Anode]
  
  \newcommand\zero{}
  \newcommand\un{}
  
  \node[Anode] {\zero}
  child {node[Bnode] {\un}
    child {node[Cnode] {\un}
      child {node[Dnode] {\zero}
        child { node[Anode] {\zero} child[pointille]}
        child { node[Cnode] {\un} child[pointille]}}
      child {node[Dnode] {\zero}
        child { node[Anode] {\zero} child[pointille]}
        child { node[Cnode] {\un} child[pointille]}}}
    child {node[Dnode] {\zero}
      child {node[Anode] {\zero}
        child { node[Bnode] {\un} child[pointille]}
        child { node[Cnode] {\un} child[pointille]}}
      child {node[Cnode] {\un}
        child { node[Dnode] {\zero}child[pointille]}
        child { node[Dnode] {\zero} child[pointille]}}}}
  child {node[Cnode] {\un}
    child {node[Dnode] {\zero}
      child {node[Anode] {\zero}
        child { node[Bnode] {\un} child[pointille]}
        child { node[Cnode] {\un} child[pointille]}}
      child {node[Cnode] {\un}
        child { node[Dnode] {\zero} child[pointille]}
        child { node[Dnode] {\zero} child[pointille]}}}
    child {node[Dnode] {\zero}
      child {node[Anode] {\zero}
        child { node[Bnode] {\un} child[pointille]}
        child { node[Cnode] {\un} child[pointille]}}
      child {node[Cnode] {\un}
        child { node[Dnode] {\zero} child[pointille]}
        child { node[Dnode] {\zero} child[pointille]}}}};
\end{tikzpicture}

      \end{tabular}
    \end{tabular}
    \caption{A Rational Balanced Tree that is not strongly balanced}
    \label{fig:ratio_bal_not_strbal}
  \end{figure}

\item \label{irratbal_not_str_bal} \emph{Irrational balanced tree that is
    not strongly balanced} -- Building an irrational tree not strongly
  balanced requires more work.  We consider a tree that which has a root
  $r$ labeled by $0$ and two children that are mechanical trees of density
  $\alpha$ and respective phases $\phi$ and $\phi+a$. We will see that
  under some conditions on $\alpha,\phi$ and $a$ this will give us an
  example of an irrational tree that is balanced but not strongly balanced
  neither rational nor Sturmian.
  
  
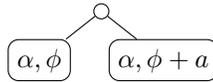
\begin{center}
    \begin{tikzpicture}[scale=0.5]
      \node[zero]{} 
      child {node[rectangle, draw, rounded corners]{$\alpha,\phi$}} 
      child {node[rectangle, draw, rounded corners] {$\alpha,\phi+a$}};
    \end{tikzpicture}
  \end{center}
  
  The two children of the root are balanced trees which means that the
  tree is balanced if and only if for all $n$:
  \begin{equation}
    \label{eq:bal_not_ratio_not_sturm}
    \lfloor (2^{n+1}-1)\alpha \rfloor \leq h_r(n+1) \leq \lfloor
    (2^{n+1}-1)\alpha \rfloor +1
  \end{equation}
  Let us call $k=\lfloor (2^n-1)\alpha+\phi\rfloor$ and
  $x=\mathrm{frac}((2^n-1)\alpha+\phi)$.
  \begin{eqnarray*}
    h_r(n+1) &=& \lfloor (2^n-1)\alpha+\phi\rfloor + \lfloor
    (2^n-1)\alpha+\phi+a\rfloor\\
    &=& k + \lfloor k+x+a \rfloor
  \end{eqnarray*}
  As $(2^{n+1}-1)\alpha = 2k+2x+\alpha-2\phi$, the equation
  \ref{eq:bal_not_ratio_not_sturm} holds if for all $x\in[0;1)$, we have:
  \begin{equation*}
    0\leq k + \lfloor k+x+a \rfloor - \lfloor  2k+2x+\alpha-2\phi \rfloor
    \leq 1
  \end{equation*}
  which holds if for all $x\in[0;1)$:
  \begin{equation*}
    0\leq \lfloor x+a \rfloor - \lfloor 2x+\alpha-2\phi \rfloor \leq 1
  \end{equation*}
  
  This equation is satisfied if and only if 
  \[ (x+a<1 \mathrm{~and~}  -1 \leq 2x-2\phi+\alpha < 1) \mathrm{~or~}
  (x+a\geq 1 \mathrm{~and~}  0 \leq 2x-2\phi+\alpha < 2)\]
  
  Looking at the extremal cases for $x+a<1$ and $x+a\geq 1$ which are
  $x=0,1-a,1$, one gets 4 relations:
  \begin{eqnarray*}
    2(1-a)-2\phi+\alpha&<&1\\
    -1&\leq&-2\phi+\alpha\\
    2-2\phi+\alpha&<&2\\
    0&\leq&2(1-a)-2\phi+\alpha.
  \end{eqnarray*}
  
  Therefore the tree is balanced if and only if
  \begin{eqnarray}
    \frac{\alpha}{2} < \phi \leq \frac{\alpha+1}{2} <
    \phi+a<1\label{eq:rela_mecha_not_strong}.
  \end{eqnarray}
  
  Moreover if $\alpha+\phi\geq 1$ and $3\alpha+\phi<2$, the tree is not
  strongly balanced since its beginning is
  
  
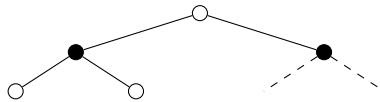
\begin{center}
    \begin{tikzpicture}[yscale=0.4]
      \node[zero]{} 
      child {node[un]{}
        child {node[zero] {}}
        child {node[zero] {}}}
      child {node[un] {}
        child[dashed] {}
        child[dashed] {}};
    \end{tikzpicture}
  \end{center}
  
  There are lots of triples $\alpha, \phi, a$ satisfying
  conditions \eqref{eq:rela_mecha_not_strong}. For example a tree with
  $\alpha=\frac{1}{3}+\epsilon$, $\phi=0.6$ and $a=0.2$ where
  $\epsilon\in\mathbb{R}\setminus\mathbb{Q}$ with $\epsilon$ small enough (for
  example $\epsilon<0.01$ works since $\frac{\alpha}{2} \approx 0.21 < 0.6
  <\frac{\alpha+1}{2} \approx 0.71 \leq 0.8 < 1$ and $\alpha+\phi>1$,
  $3\alpha+\phi\approx 1.9 < 2$).
\end{enumerate}

\bibliographystyle{plain}
\bibliography{localBiblio}

\end{document}